\documentclass[final,5p,times,twocolumn,authoryear]{elsarticle}

\usepackage{amssymb}
\usepackage{amsthm}
\usepackage{amsfonts}
\usepackage{amsmath}  
\usepackage{amssymb}  
\usepackage[colorlinks = true, urlcolor  = blue]{hyperref}  
\usepackage{tabularx}  
\usepackage{pdflscape}  
\usepackage{changepage}  
\usepackage{booktabs}  
\usepackage{lineno}
\usepackage{caption}
\usepackage[normalem]{ulem}

\journal{Icarus}

\begin{document}

\begin{frontmatter}



\title{Decameter-sized Earth impactors -- I: Orbital properties}


\author[pa, ws]{Ian Chow\corref{coric}}
\cortext[coric]{Corresponding author.  \\ E-mail address: \href{mailto:ichow9@uwo.ca}{ichow9@uwo.ca} (I. Chow)}

\affiliation[pa]{organization={Department of Physics and Astronomy, University of Western Ontario},
            addressline={1151 Richmond St}, 
            city={London},
            postcode={N6A 3K7}, 
            state={Ontario},
            country={Canada}}

\author[pa,ws]{Peter G. Brown}

\affiliation[ws]{organization={Western Institute for Earth and Space Exploration, University of Western Ontario},
            addressline={Perth Drive}, 
            city={London},
            postcode={N6A 5B7}, 
            state={Ontario},
            country={Canada}}

\begin{abstract}
Numerous decameter-sized asteroids have been observed impacting Earth as fireballs. These objects can have impact energies equivalent to hundreds of kilotons of TNT, posing a hazard if they impact populated areas. Previous estimates of meteoroid flux using fireball observations have shown an Earth impact rate for decameter-size objects of about once every $2$-$3$ years. In contrast, telescopic estimates of the near-Earth asteroid population predict the impact rate of such objects to be of order $20$-$40$ years, an order-of-magnitude difference. While the cause of this discrepancy remains unclear, tidal disruption of a larger near-Earth body has been proposed as an explanation for these excess decameter-sized impactors. The release in 2022 of previously classified United States Government (USG) satellite sensor data for fireball events has provided a wealth of new information on many of these impacts. Using this newly available USG sensor data, we present the first population-level study characterizing the orbital and dynamical properties of 14 decameter-sized Earth impactors detected by USG sensors since 1994, with a particular focus on searching for evidence of tidal disruption as the cause of the impact rate discrepancy. We find there is no evidence for recent ($\lesssim 10^4$ years) tidal disruption and weak evidence for longer-term tidal disruption in the decameter impactor population, but that the latter conclusion is limited by small number statistics. We also investigate the origins of both the impactor and near-Earth asteroid populations of decameter-sized objects from the main asteroid belt. We find that both populations generally originate from the same source regions: primarily from the $\nu_6$ secular resonance ($\sim70$\%) with small contributions from the Hungaria group ($\sim20$\%) and the 3:1 Jupiter mean-motion resonance ($\sim10$\%).

\end{abstract}




\end{frontmatter}



\section{Introduction}

The population and physical characteristics of decameter-sized near-Earth objects (NEOs) is poorly known. 
This size range is at the lower end of detection for telescopic surveys while also being rare as Earth impactors.  
This paucity of physical and orbital data is reflected in the small number ($\sim$300) of well-determined decameter and smaller NEO orbits. 
It is further demonstrated by the fact that only $\sim$10 near-Earth asteroids (NEAs) in this size range have known rotation periods \citep{bolin_rotation_2024}, 
the small number ($\sim$50) of such NEAs with spectroscopic data \citep{marsset_debiased_2022}
and the handful of bulk density estimates for decameter NEAs \citep{mommert_physical_2014, mommert_constraining_2014}.

Despite this dearth of physical information, decameter NEOs are of interest in near-Earth asteroid mining, which has been explored as a cheaper and more environmentally sustainable option \citep{hein_exploring_2018, fleming_mining_2023} to mining on Earth. Previous studies evaluating potential targets for asteroid mining \citep{garcia_yarnoz_easily_2013, xie_target_2021} have identified a number of suitable candidates of decameter size. They also represent prime targets for exploration as many have very low $\Delta$v with respect to Earth \citep{elvis_ultra-low_2011}. 
Decameter sized NEAs are also of interest as they are part of the NEA population which are the immediate parent asteroids of meteorites \citep{borovicka_are_2015}. 

The physical characteristics for some decameter NEAs have been inferred indirectly from telescopic measurements through comparison to models \citep{sanchez_strength_2014}. However, the internal structure of decameter NEAs remains poorly constrained from telescopic data alone.

Fireball observations of decameter-size impactors afford a unique window into their internal structure while also providing orbital context, assuming sufficient measurements are available. 
Decameter-size objects therefore represent a unique population, as sufficient data from both telescopic and fireball observations at this size regime allows for population-level inferences.
The numerous studies of the Chelyabinsk bolide \citep{brown_500-kiloton_2013, popova_chelyabinsk_2013, borovicka_trajectory_2013} 
demonstrate
what is possible when modeling and measurements of large bolides are combined. 
These works showed that the Chelyabinsk meteoroid was a heterogeneous object composed of a small mass fraction ($\sim$1\%) of monolithic meter-sized rocks held together by a globally weak (strength of order $0.7$ MPa) grain/cemented structure. The associated impact blast wave injured $1600$ people and thousands of windows were broken in Chelyabinsk and the surrounding area \citep{kartashova_study_2018}.

While Chelyabinsk remains by far the best-documented decameter-class impactor, over the past $30$ years, numerous asteroids of decameter size (which we define here as larger than approximately $7.5$ meters in diameter given uncertainties in connecting mass to size) have been observed as fireballs impacting Earth's atmosphere 
\citep{tagliaferri_analysis_1995, arrowsmith_global_2008, silber_infrasonic_2011, borovicka_trajectory_2013}.
These decameter-size objects can have equivalent impact energies of up to hundreds of kilotons of TNT \citep[$1$ kT TNT $= 4.184 \times 10^{12}$ J;][]{brown_500-kiloton_2013}, posing a serious hazard if they impact populated areas. Indeed, decameter-class impactors are the most likely to cause significant ground damage in the near future \citep{boslough_updated_2015, gi_frequency_2018}.

\begin{figure*}
    \centering
    \includegraphics[width=1.\textwidth]{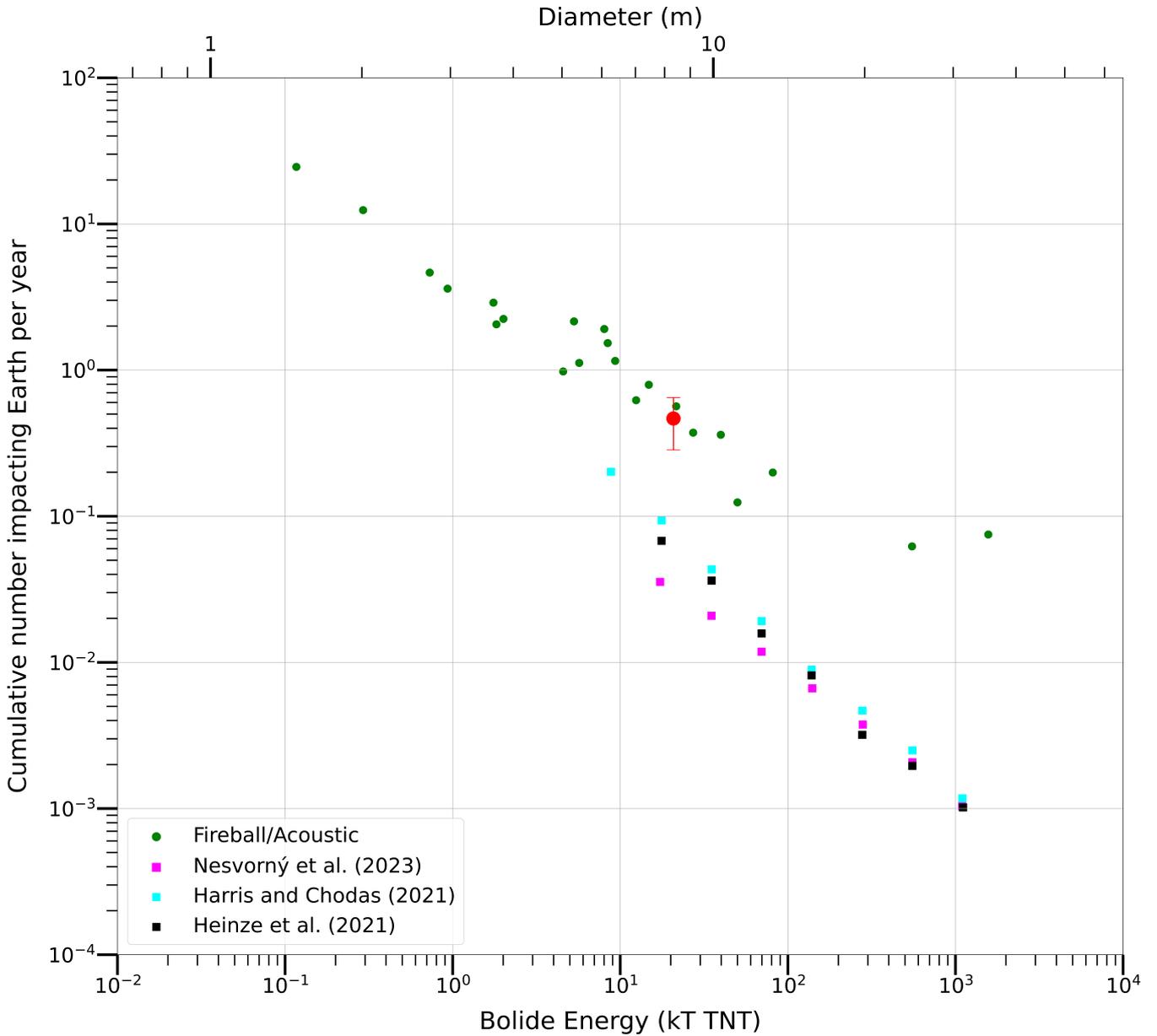}
    \caption{The size-impact energy flux distribution of Earth-impacting objects and telescopic populations of near-Earth asteroids (NEAs). This work's estimate of the decameter impact rate of 0.467 impacts per year using United States Government (USG) sensor data from NASA's Center for Near Earth Object Studies (CNEOS) database of individual events is shown as the red dot. The error bars reflect the $1\sigma$ uncertainty assuming impacts are modeled as a Poisson process. Note that for the fireball flux, the impact rate is directly observed together with energy. The diameter (top axis) is inferred by assuming a mean bulk density of $\rho = 1500$ kg/m$^{3}$. Previous estimates of impact flux at different sizes using fireball and acoustic data by \citet{brown_flux_2002}, \citet{silber_estimate_2009}, and \citet{brown_500-kiloton_2013} are plotted in green. Also shown are the inferred impact rates from  
    three NEO population models based on $5$ recent telescopic surveys \citep{heinze_neo_2021, harris_population_2021, nesvorny_neomod_2023}, assuming a geometric albedo of $p_v = 0.18$, annual impact probability of $f_{\mathrm{imp}} = 2.6 \times 10^{-9}$ yr$^{-1}$, impact velocity of $v = 20$ km/s and bulk density of $\rho = 1500$ kg/m$^{3}$ for each NEA.
    The annual impact probability is based on the value estimated by \citet{nesvorny_neomod_2023} for NEAs with absolute magnitude $H = 28$, corresponding to objects of size $\sim8$m. The impact velocity is based on previous estimates for average NEA impact velocity ranging from $18-21$ km/s \citep{brown_flux_2002, tricarico_near-earth_2017, harris_population_2021}. 
    The order-of-magnitude discrepancy between the observed fireball/acoustic impact rates and the telescopically inferred impact rates is visible in the decameter size regime.}
    \label{fig:decameter_gap}
\end{figure*}

Recently, sufficient number statistics for both telescopic surveys of NEOs and fireball observations of Earth impacts in the decameter size regime has allowed for direct comparison of the apparent flux of telescopic and fireball observations \citep{nesvorny_neomod_2024}. Previous estimates of near-Earth meteoroid flux using fireball and infrasonic data 
\citep{brown_flux_2002, silber_estimate_2009, brown_500-kiloton_2013} have generally converged on an Earth impact rate for decameter objects of roughly once every $2$-$3$ years. In contrast, several independent telescopic population estimates place the decameter impactor rate at Earth at roughly once every $20$-$40$ years \citep{heinze_neo_2021, harris_population_2021, nesvorny_neomod_2023}, an order-of-magnitude difference from fireball estimates.
Figure \ref{fig:decameter_gap} highlights this discrepancy between the observed impact rate based on fireball data and the inferred impact rate from telescopic surveys. The cause of this ``decameter gap" remains unclear.

One proposed explanation for this discrepancy is that the excess impact flux in the decameter size regime is the product of recent tidal disruption of one or more larger near-Earth bodies \citep{granvik_tidal_2024, nesvorny_neomod_2024}. 
In this interpretation, a small subset of NEAs originating from a tidal disruption would dominate the impactor population. These NEAs would have much higher impact probability than the ``background" population, meaning the average impact probability assumed to convert the telescopic NEA population numbers to an inferred impact rate would be too low. This would effectively account for the gap by shifting the telescopic curves upward in Fig. \ref{fig:decameter_gap}.

Here we explore this possibility, focusing on the orbital distribution of observed decameter (and smaller) impactors to search for a possible tidal disruption signature.
Our work is motivated in part by the release in $2022$ by the United States Space Force of decades of previously classified United States Government (USG) satellite sensor data for fireball events\footnote{\href{https://cneos.jpl.nasa.gov/fireballs/}{cneos.jpl.nasa.gov/fireballs/}}. These new lightcurves and associated orbital data allow for a more detailed analysis of many large events. In particular, we identify $14$ impacts of decameter-sized objects over the past $30$ years from these USG sensor data, an impact rate consistent with previous fireball estimates yet still much higher than telescopic estimates. The size of these impactors was estimated based on the reported speed and energy from USG sensors combined with a notional bulk density of $1500$ kg/m$^{3}$. The choices of these values and their associated uncertainties are discussed later.

While systematic uncertainties in fireball energy or albedo could contribute to the ``decameter gap", they are unlikely to be a significant factor. \citet{devillepoix_observation_2019} analyzed systematic uncertainties in USG sensor-recorded fireball energies and independently validated them against acoustic measurements and ground-to-satellite calibration. More recently, USG energies have been compared to common fireballs detected by the Geostationary Lightning Mapper (GLM) instrument onboard the Geostationary Operational Environmental Satellite (GOES) weather satellites \citep{wisniewski_determining_2024}. The reported USG energies have been found to be generally accurate, with \citet{wisniewski_determining_2024} finding agreement to much better than a factor of two for well observed common fireballs.
Moreover, as previous studies of NEAs have not identified any size-dependence of albedo down to roughly decameter size \citep[][]{mainzer_characterizing_2012, granvik_super-catastrophic_2016, devogele_visible_2019}, albedo is unlikely to be a significant source of systematic uncertainty either.
As such, we focus on assessing tidal disruption as the primary explanation for the discrepancy in the observed impact flux.

This paper is structured as follows:
Section \ref{sec:data} describes the data used. 
In Section \ref{sec:uncertainty_estimation}, we outline our methodology for estimating uncertainty in the orbits of the $14$ decameter impactors recorded by USG sensors. 
In Section \ref{sec:orbital_similarity_ks_test}, we compare the $3$-D $\left(a, e, i\right)$ distributions of the decameter impactors and telescopically observed NEAs to determine whether the observed decameter impactor population is representative of the true decameter NEA population. 
In Section \ref{sec:orbital_similarity_d_criteria}, we evaluate the recent tidal disruption hypothesis for the large discrepancy between the telescopically inferred impact rate and the observed impact rate, as shown in Fig. \ref{fig:decameter_gap}. We also explore the possibility that the decameter impact enhancement could be the result of longer-term tidal disruption in Section \ref{sec:long_term_td}.
In Section \ref{sec:source_regions} we analyze the dynamical origins of the decameter population, including both impactors and telescopically observed objects, and determine what source regions of the main asteroid belt they originate from. 
We discuss our results and possible avenues for further research in section \ref{sec:discussion}.
Finally, section \ref{sec:conclusions} summarizes our conclusions.

\begin{table*}[t]
    \begin{center} 
    \caption{Characteristics and state vectors of decameter Earth impactors recorded by U.S. Government satellites.}
    \begin{tabularx}{\textwidth}{l c c c c c c c c c}
    \toprule\toprule
       UTC Date & Lat. & Lon. & $\phi$ & $\theta$ & Height & Velocity & Impact Energy & Diameter & Ref. \\
       (YYYY-MM-DD) & $\left(^\circ\mathrm{N}\right)$ & $\left(^\circ\mathrm{E}\right)$ & $\left(^\circ\right)$ & $\left(^\circ\right)$ & (km) & (km/s) & (kT TNT) & (m) & \\
       \hline
       1994-02-01 & $2.7$ & $164.1$ & $119.6$ & $44.5$ & $21.0$ & $24.0$ & $30.0$ & $8.34$ & (1) \\
       1999-01-14 & $-44.0$ & $-129.6$ & $237.6$ & $35.4$ & $32.0$ & $15.2$ & $9.8$ & $8.31$ & (2) \\
       2004-09-03 & $-67.7$ & $18.2$ & $262.05$ & $48.14$ & $31.5$ & $13.0$ & $13.0$ & $9.50$ & (3) \\
       2004-10-07 & $-27.3$ & $71.5$ & $240.13$ & $62.79$ & $35.0$ & $19.2$ & $18$ & $8.16$ & (4) \\ 
       2006-12-09 & $26.2$ & $26.0$ & $98.71$ & $85.51$ & $26.5$ & $15.9$ & $14.0$ & $8.51$ & \\
       2009-10-08 & $-4.2$ & $120.6$ & $27.46$ & $22.54$ & $19.1$ & $19.2$ & $33.0$ & $9.99$ & (5) \\
       2010-07-06 & $-34.1$ & $-174.5$ & $50.99$ & $46.13$ & $26.0$ & $15.7$ & $14.0$ & $8.59$ & \\
       2010-12-25 & $38.0$ & $158.0$ & $147.20$ & $29.11$ & $26.0$ & $18.5$ & $33.0$ & $10.39$ & \\
       2013-02-15 & $54.8$ & $61.1$ & $99.90$ & $74.08$ & $23.3$ & $18.6$ & $440$ & $24.20$ & (6) \\
       2013-04-30 & $35.5$ & $-30.7$ & $297.77$ & $50.50$ & $21.2$ & $12.1$ & $10.0$ & $9.13$ & \\
       2016-02-06 & $-30.4$ & $-25.5$ & $260.49$ & $68.09$ & $31.0$ & $15.6$ & $13.0$ & $8.41$ & \\
       2018-12-18 & $56.9$ & $172.4$ & $349.43$ & $21.41$ & $26.0$ & $25.0$ & $49.0$ & $9.56$ & (7) \\
       2020-12-22 & $31.9$ & $96.2$ & $351.80$ & $85.08$ & $35.5$ & $13.7$ & $9.5$ & $8.30$ & \\
       2022-02-07 & $-28.7$ & $11.4$ & $27.41$ & $63.32$ & $26.5$ & $13.1$ & $7.0$ & $7.69$ & \\
       \bottomrule
    \end{tabularx}\label{tab:decameter_impactors}
    \end{center}
    \begin{small}
    \textbf{Note:} Negative values of latitude and longitude represent south and west, respectively. $\phi$ and $\theta$ are the azimuth and zenith angle of the meteor radiant. 
    The height and velocity are given at the height of peak brightness.
    Object diameter is estimated from velocity and impact energy assuming a spherical shape with a bulk density of $1500$ kg/m$^{3}$.
    The data are obtained from NASA's Center for Near Earth Object Studies (CNEOS) database (\href{https://cneos.jpl.nasa.gov/fireballs/}{cneos.jpl.nasa.gov/fireballs/}), which does not provide formal uncertainties. We refer the reader to previous studies evaluating the general accuracy of CNEOS measurements \citep{devillepoix_observation_2019, brown_proposed_2023} for further discussion regarding uncertainties.
    In cases where the CNEOS data has been supplemented with additional literature sources, a corresponding reference is provided.
    
    \textbf{References:} (1) \citet{tagliaferri_analysis_1995}; (2) \citet{pack_recent_1999}; (3) \citet{klekociuk_meteoritic_2005}; (4) \citet{arrowsmith_global_2008}; (5) \citet{silber_infrasonic_2011}; (6) \citet{borovicka_trajectory_2013, brown_500-kiloton_2013}; (7) Retrieved from the CNEOS database at December 15, 2021 \citep{borovicka_satellite_2020}.
    \end{small}
\end{table*}

\section{Data}\label{sec:data}
The basic data for the decameter impactors were obtained largely from NASA's Center for Near Earth Object Studies (CNEOS) Fireball and Bolide Database, which reports fireball events observed by USG satellites. In some cases this information was supplemented with literature sources.

The CNEOS data contains velocity information as well as height of peak brightness for the associated fireball and its estimated total energy. Detailed lightcurves of bolide brightness as a function of time are also available.\footnote{\href{https://cneos.jpl.nasa.gov/fireballs/lc/}{cneos.jpl.nasa.gov/fireballs/lc/}}

We compute each object's diameter from the reported velocity and impact energy assuming a spherical shape with a bulk density of $1500$ kg/m$^3$, which we adopt as a value in reasonable quantitative agreement with other estimates. Our adopted value is based on the
estimated bulk densities for the decameter-sized NEAs 2009 BD \citep{mommert_constraining_2014} of $1700_{-400}^{+700}$ kg/m$^3$ and 2011 MD \citep{mommert_physical_2014} of $1100_{-500}^{+700}$ kg/m$^3$ determined using thermophysical and non-gravitational force modeling, as well as the bulk density of $2300\pm200$ kg/m$^3$ inferred for 2008 TC3 \citep{jenniskens_impact_2009}, the only meteorite-producing fireball for which telescopic data are sufficient to constrain the shape and size of the impactor. 

Our bulk density choice is also supported by the lower average bulk density of NEAs compared to main-belt asteroids \citep{carry_density_2012} and is similar to the in-situ measured bulk densities for the LL-chondrite-like NEA Itokawa of $1950 \pm 140$ kg/m$^3$\citep{abe_mass_2006}, the primitive NEAs Bennu of $1190\pm13$ kg/m$^3$ \citep{lauretta_unexpected_2019} and Ryugu of $1190 \pm 20$ kg/m$^3$ \citep{watanabe_hayabusa2_2019}, and the Didymos/Dimorphos system of the Double Asteroid Redirection Test (DART) of $2170 \pm 350$ kg/m$^3$ \citep{rivkin_double_2021}. Finally, our adopted bulk density is also comparable to recent estimates of numerous NEA bulk densities from \textit{Gaia} data \citep{dziadura_yarkovsky_2023}, which showed a modal density for S-type asteroids of $\sim 1500$ kg/m$^3$.
We recognize that while there may be some spread in the bulk density of the decameter impactors, there is no way of quantifying this uncertainty. However, as bulk density is proportional only to the one-third power of diameter, it is unlikely to be a significant factor in the ``decameter gap."

Here we define decameter-sized objects as those with a diameter above $7.5$ meters. The upper bound is the approximate size ($\sim20$m) of the 2013 Chelyabinsk fireball \citep{borovicka_trajectory_2013}, the largest and most energetic impactor for which fireball data is available. 

From the CNEOS database, we identify $14$ decameter-sized Earth impactors over the past $30$ years (from February 1, 1994 to February 1, 2024) using the reported speed and energy to estimate mass and our adopted bulk density of $1500$ kg/m$^3$ to derive diameter. These $14$ impactors are summarized in Table \ref{tab:decameter_impactors}.

The population of telescopically-observed decameter objects were selected using NASA's Small-Body Database.\footnote{\href{https://ssd.jpl.nasa.gov/tools/sbdb\_query.html}{ssd.jpl.nasa.gov/tools/sbdb\_query.html}} The asteroid diameter $D$ in kilometers was estimated from the geometric albedo $p_v$ and absolute magnitude $H$ using the following equation \citep{harris_revision_1997}:
\begin{equation}
    D = 10^{3.1236 - 0.5\log_{10}\left(p_v\right) - 0.2H}\label{eqn:asteroid_size}
\end{equation}
We assumed a geometric albedo of $p_v = 0.18$ for all objects based on the recommended value by \citet{nesvorny_neomod_2024-1} when converting $H$ to size for objects fainter than $H > 22$, and limited the absolute magnitude to $26.5 \leq H \leq 28$ for the decameter population. This corresponds to a diameter of $16$m $\gtrsim D \gtrsim 7.9$m from equation \ref{eqn:asteroid_size}, similar to the bounds for our impactor sample. Finally, we only consider objects with an uncertainty parameter (also called condition code) of $U \leq 5$, to filter out objects with poorly defined orbits. This process yields a sample of $295$ telescopically observed decameter-sized NEAs whose orbits are relatively well-defined.
\begin{figure*}[t]
    \centering
    \includegraphics[width=1.\textwidth]{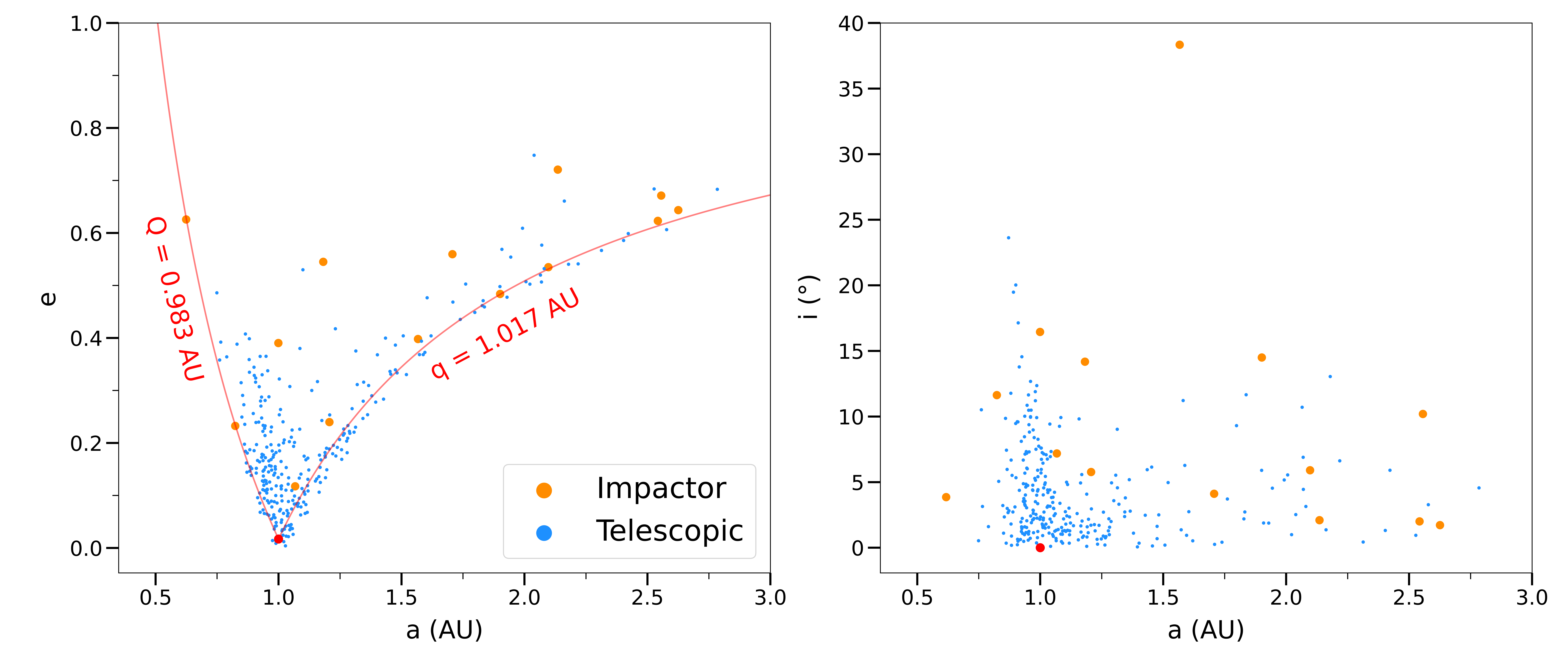}
    \caption{The semi-major axes $a$ of all $14$ decameter-sized Earth impactors (orange) and $295$ telescopically well-observed decameter-sized NEAs (blue) that we identify as described in the main text, as a function of their eccentricities $e$ (left) and inclinations $i$ (right). The red dot represents the Earth's position in $a$-$e$ and $a$-$i$ parameter space. The red lines on the $a$-$e$ plot are the set of all orbits with aphelion distance $Q = 0.983$ AU or perihelion distance $q = 1.017$ AU, representing the minimum aphelion and maximum perihelion distance required for an object to collide with the Earth.
    }
    \label{fig:aei_telescopic_vs_impactors}
\end{figure*}

Fig. \ref{fig:aei_telescopic_vs_impactors} shows the semi-major axes $a$, eccentricities $e$ and inclinations $i$ of all $14$ decameter-sized Earth impactors and $295$ telescopically observed decameter-sized NEAs that we have identified, compared to the Earth's orbit. Table \ref{tab:nominal_orbits} shows the nominal orbits for all 14 impactors together with the Tisserand parameter with respect to Jupiter, $T_J$, for each object. The observed impactor population differs from the raw observed NEA population in that there is a higher proportion of objects with large semi-major axes and eccentricities, with several on the border between asteroidal and cometary orbits ($T_J \sim$ 3). The uncertainty in these orbits will be discussed in the next section.  

\begin{landscape}
\thispagestyle{empty}  
\captionsetup{labelfont=bf,
              justification=raggedright,
              singlelinecheck=off}
\begin{table*}[t]
    \caption{Nominal pre-impact orbits of decameter Earth impactors}
    \centering
    \begin{adjustwidth}{-7.5cm}{}
    \begin{tabular}{l c c c c c c c c c c}
        \toprule\toprule
        UTC Date & $a$ & $e$ & $i$ & $q$ & $\omega$ & $\Omega$ & $T_J$ & $Q$ & $f_\mathrm{imp}$ & 
        Clones with \\ 
        & & & & & & & & & & $10~\times~$avg. $f_\mathrm{imp}$ \\
        (YYYY-MM-DD) & (AU) & & $\left(^\circ\right)$ & (AU) & $\left(^\circ\right)$ & $\left(^\circ\right)$ & & (AU) & ($10^{-9}$ yr$^{-1}$) & ($\%$) \\
        \bottomrule
        1994-02-01 & 2.14$\pm 0.69$ & 0.72$\pm 0.083$ & 2.10$\pm 5.41$ & 0.60$\pm 0.092$ & 273.34$\pm 56.75$ & 132.89$\pm 0.028$ & 3.32$\pm 0.83$ & 3.67$\pm 1.37$ & $2.63$ & $2.8$  \\
        1999-01-14 & 1.90$\pm 0.75$ & 0.48$\pm 0.14$ & 14.50$\pm 4.22$ & 0.98$\pm 0.030$ & 352.80$\pm 58.52$ & 113.71$\pm 0.0032$ & 3.76$\pm 0.78$ & 2.82$\pm 1.49$ & $5.44$ & $0.5$ \\
        2004-09-03 & 0.82$\pm 0.10$ & 0.23$\pm 0.092$ & 11.63$\pm 5.77$ & 0.63$\pm 0.11$ & 191.97$\pm 22.58$ & 341.25$\pm 0.0054$ & 7.08$\pm 0.53$ & 1.02$\pm 0.14$ & $11.35$ & $2.2$ \\
        2004-10-07 & 2.56$\pm 1.15$ & 0.67$\pm 0.14$ & 10.20$\pm 2.16$ & 0.84$\pm 0.021$ & 307.16$\pm 60.57$ & 14.55$\pm 2.1\times 10^{-4}$ & 3.06$\pm 0.79$ & 4.27$\pm 2.30$ & $0.786$ & $0.1$  \\
        2006-12-09 & 2.63$\pm 0.93$ & 0.64$\pm 0.12$ & 1.73$\pm 2.92$ & 0.94$\pm 0.11$ & 150.77$\pm 69.70$ & 256.95$\pm 0.013$ & 3.07$\pm 0.91$ & 4.31$\pm 1.83$ & $6.77$ & $25.2$ \\
        2009-10-08 & 1.18$\pm 0.24$ & 0.55$\pm 0.09$ & 14.18$\pm 4.10$ & 0.54$\pm 0.066$ & 71.97$\pm 92.20$ & 194.83$\pm 4.6\times 10^{-4}$ & 5.18$\pm 0.63$ & 1.83$\pm 0.46$ & $1.49$ & $1.7$ \\
        2010-07-06 & 2.54$\pm 1.49$ & 0.62$\pm 0.16$ & 2.01$\pm 2.73$ & 0.96$\pm 0.060$ & 328.30$\pm 65.31$ & 284.56$\pm 0.13$ & 3.14$\pm 0.92$ & 4.13$\pm 2.98$ & $7.34$ & $23.6$ \\
        2010-12-25 & 1.00$\pm 0.12$ & 0.39$\pm 0.11$ & 16.45$\pm 4.82$ & 0.61$\pm 0.091$ & 69.15$\pm 12.91$ & 273.91$\pm 9.8 \times 10^{-4}$ & 5.98$\pm 0.53$ & 1.39$\pm 0.23$ & $2.29$ & $3.1$ \\
        2013-02-15 & 1.71$\pm 0.020$ & 0.56$\pm 0.0048$ & 4.12$\pm 0.094$ & 0.75$\pm 0.010$ & 109.50$\pm 7.91\times 10^{-4}$ & 326.44$\pm 0.13$ & 4.00$\pm 0.023$ & 2.66$\pm 0.026$ & $2.70$ & $0.0$ \\
        2013-04-30 & 1.07$\pm 0.21$ & 0.12$\pm 0.11$ & 7.19$\pm 5.05$ & 0.94$\pm 0.061$ & 247.51$\pm 28.87$ & 40.01$\pm 0.0088$ & 5.77$\pm 0.56$ & 1.19$\pm 0.41$ & $6.94$ & $1.3$ \\
        2016-02-06 & 0.62$\pm 0.066$ & 0.60$\pm 0.12$ & 3.86$\pm 6.50$ & 0.25$\pm 0.12$ & 2.80$\pm 45.98$ & 316.97$\pm 0.0073$ & 8.98$\pm 0.67$ & 0.99$\pm 0.030$ & $53.74$ & $85.0$ \\
        2018-12-18 & 1.57$\pm 0.45$ & 0.40$\pm 0.13$ & 38.34$\pm 3.76$ & 0.94$\pm 0.051$ & 211.22$\pm 13.36$ & 266.75$\pm 1.7 \times 10^{-6}$ & 4.11$\pm 0.69$ & 2.19$\pm 0.90$ & $3.22$ & $1.6$ \\
        2020-12-22 & 2.10$\pm 0.67$ & 0.53$\pm 0.13$ & 5.91$\pm 5.48$ & 0.98$\pm 0.053$ & 192.62$\pm 14.17$ & 271.29$\pm 0.0091$ & 3.55$\pm 0.78$ & 3.22$\pm 1.32$ & $5.14$ & $8.8$ \\
        2022-02-07 & 1.21$\pm 0.40$ & 0.24$\pm 0.14$ & 5.77$\pm 1.67$ & 0.92$\pm 0.050$ & 230.25$\pm 30.66$ & 318.73$\pm 0.0015$ & 5.24$\pm 0.71$ & 1.50$\pm 0.80$ & $5.22$ & $2.3$ \\
        \bottomrule
        \textit{RMS Error} & $0.68$ & $0.12$ & $4.27$ & $0.074$ & $51.33$ & $0.051$ & $0.70$ & $1.34$ \\
        \bottomrule
    \end{tabular}
    \begin{small}
    The nominal pre-impact orbits of all $14$ decameter-size Earth impactors recorded by USG sensors from 1994-2024.
    The orbital elements $a$, $e$, $i$, $q$, $\omega$, $\Omega$ are the semi-major axis, eccentricity, inclination, perihelion distance, argument of perihelion, and longitude of ascending node, respectively.
    Here the angular elements are in J2000.0.
    $T_J$ is the Tisserand parameter with respect to Jupiter. 
    The uncertainties for each object are given as the $1\sigma$ standard deviation of the $1000$ Monte Carlo clones generated using the procedure described in Section \ref{sec:uncertainty_estimation}. 
    Also shown is the nominal orbital impact probability with the Earth as computed using the method of \citet{pokorny_opik-type_2013}, and the percentage of $1000$ clones per event which have $10$ times the typical NEA impact probability of $f_\mathrm{imp} = 2.6 \times 10^{-9}$ yr$^{-1}$.
    The orbits of all $14$ decameter impactors are identified as likely classical asteroidal based on the criteria outlined by \citet{borovicka_data_2022} (aphelion distance $Q < 4.9$ AU).
    \end{small}
     \end{adjustwidth}
    \label{tab:nominal_orbits}
\end{table*}
\end{landscape}

\section{Impactor orbital uncertainty estimation}\label{sec:uncertainty_estimation}

The velocity vectors for USG fireballs do not have a reported uncertainty. The accuracy of the associated orbits on a per object basis is unknown. There is substantial debate in the community whether the uncertainties are quite large on average \citep{devillepoix_observation_2019, brown_proposed_2023} or more quantifiable/certain \citep{pena-asensio_orbital_2022, socas-navarro_how_2024}. However, in the absence of any specific information on a per event basis this is speculative, and strong statements or conclusions about the specific orbit for any given fireball recorded solely by USG sensors should certainly be treated with great caution.  

One exception is the pre-impact orbit of the 2013 Chelyabinsk bolide. As it is extremely well-characterized compared to the other $13$ decameter impactors, here we estimate its orbital uncertainty by drawing $1000$ Monte Carlo clones from a Gaussian distribution using the orbital elements and corresponding uncertainties given in Table $2$ of \citet{borovicka_trajectory_2013}.

However, uncertainties for the orbital elements of the other decameter impactors recorded by USG sensors are not available, as they were only recorded by USG sensors.
As such, we estimate the uncertainty in the orbits of the $13$ other decameter impactors from the USG sensor data by instead using a calibration set of $18$ fireball events that were observed both by USG sensors and separately by ground-based instruments. 
We note that these events are much less energetic than our decameter impactors and as such we expect these uncertainties to be conservative -- i.e. our larger (brighter) fireballs likely have lower uncertainties. 

The calibration set consists of the $17$ fireballs listed in Table $3$ of \citet{brown_proposed_2023} as well as the May 20, 2023 fireball observed over Queensland, Australia by cameras of the Global Meteor Network (Vida, personal communication, 2023).
\citet{devillepoix_observation_2019} identified a large discrepancy between the velocity vector recorded by USG sensors and ground-based observations for the fireball $2008$ TC$3$/Almahatta Sitta. This discrepancy was subsequently analyzed by \citet{pena-asensio_orbital_2022}, who argued that it is the result of a typographical error in the USG values reported on the CNEOS website; similar errors have been found (and corrected) on the website before, so this is a reasonable explanation. Indeed, \citet{pena-asensio_orbital_2022} found much better agreement with ground-based observations when reversing the sign of the $z$-component of the CNEOS velocity vector from $3.8$ km/s to $-3.8$ km/s. As such, we use the reversed $z$-component velocity vector suggested by \citet{pena-asensio_orbital_2022} instead of the CNEOS reported value in our work as well.

\begin{figure*}[t]
\centering
\begin{minipage}{.495\textwidth}
  \centering
  \includegraphics[width=1.\linewidth]{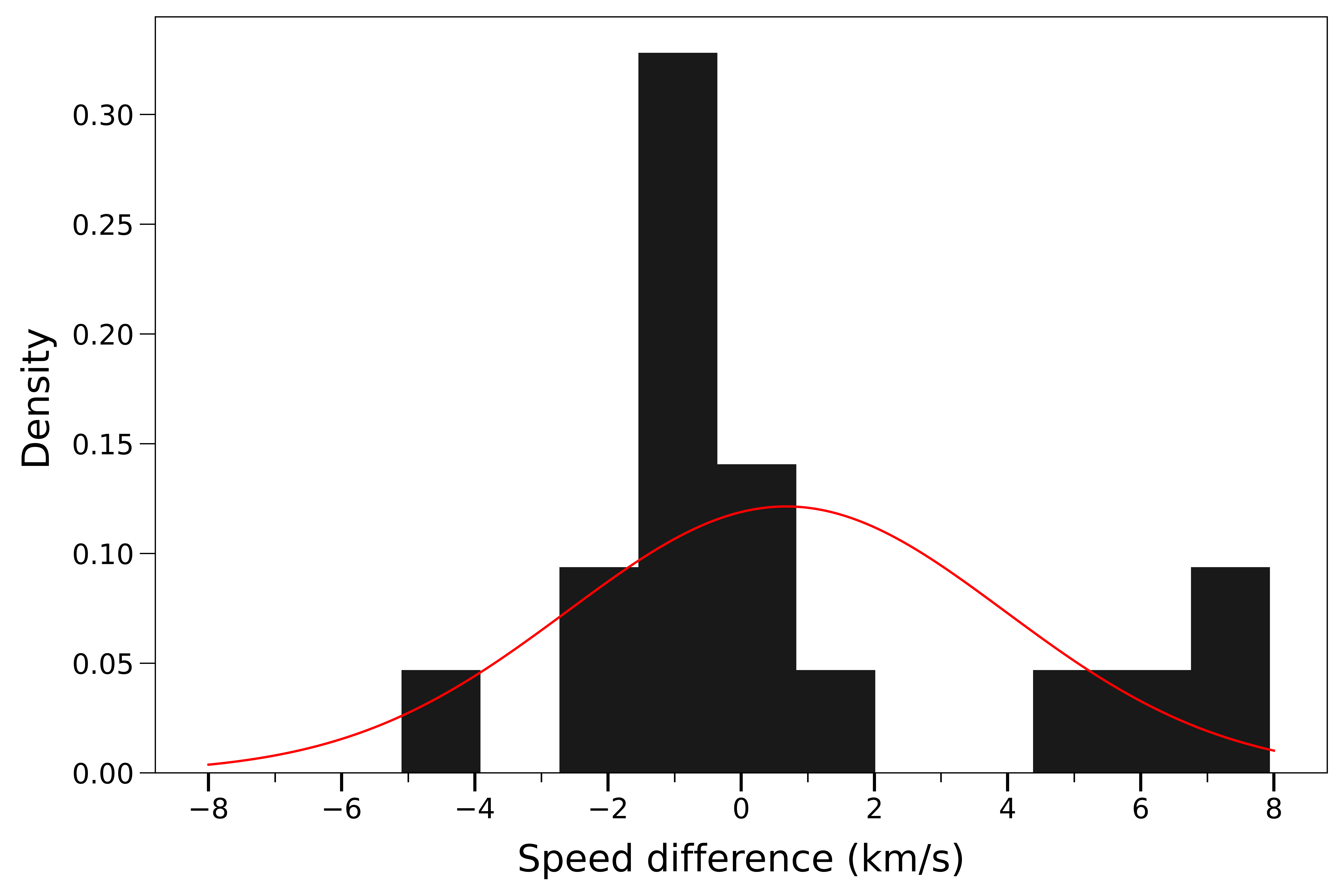}
\end{minipage}
\begin{minipage}{.495\textwidth}
  \centering
  \includegraphics[width=1.\linewidth]{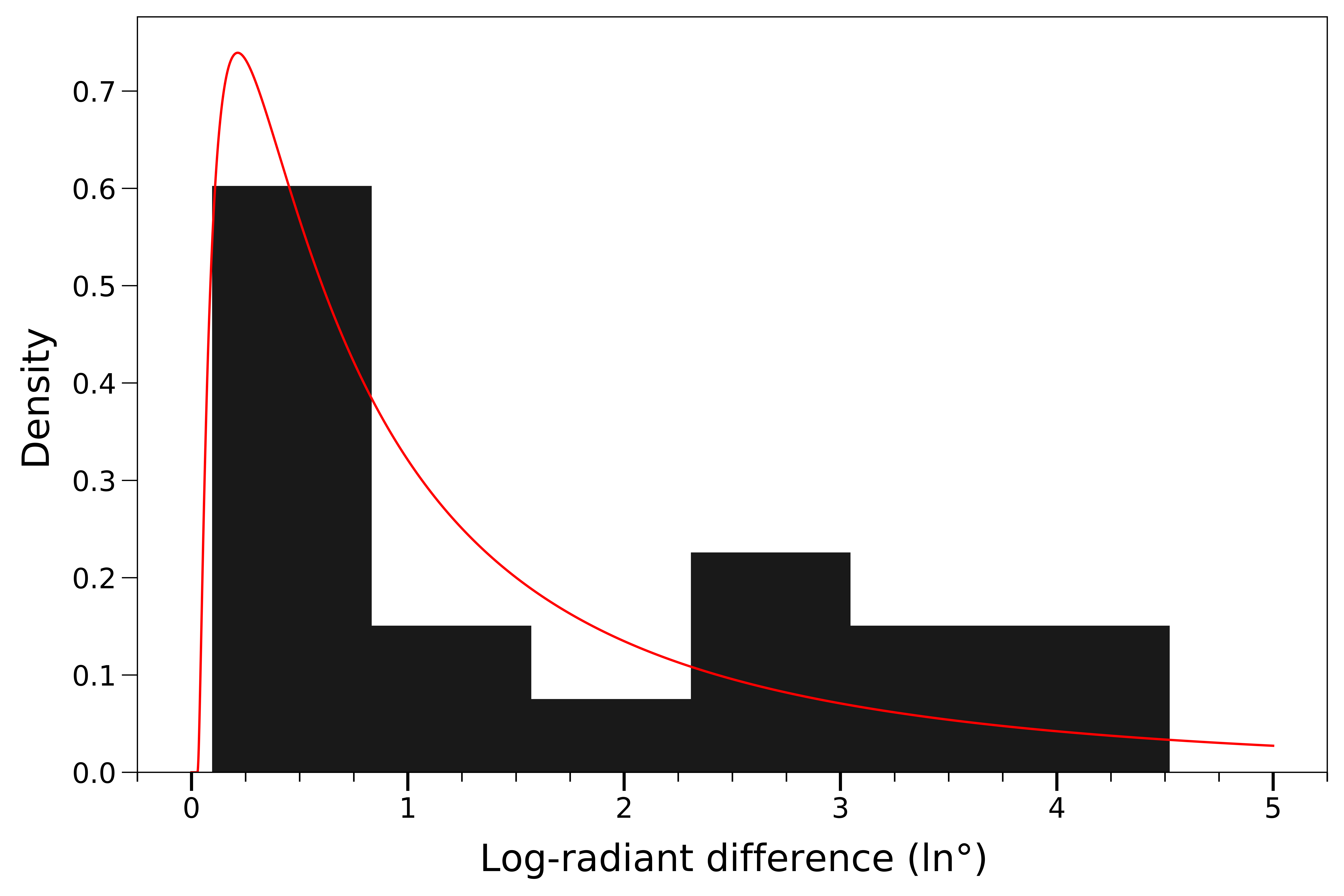}
\end{minipage}
\caption{Estimated probability distributions and corresponding histograms for speed uncertainty (left) and radiant natural log-uncertainty (right), computed using a set of $18$ calibrated fireballs identified by \citet{brown_proposed_2023} that were observed both by USG sensors and by ground-based instruments. 
The speed difference is measured at the time of peak brightness, while the radiant difference is computed as the angular separation between the reported radiants.
The speed uncertainty is modeled as a Gaussian with $\mu = 0.67$, $\sigma = 3.28$ and the log-radiant uncertainty is estimated as a log-Gaussian with $\mu = 0.96, \sigma = 1.28$. These are the respective best-fit solutions to the data. 
}
\label{fig:uncertainty_histograms}
\end{figure*}

For each fireball in the calibration set, the difference in the fireball's speed as well as its radiant as measured by USG sensors and by ground-based observations is computed and taken to be the uncertainty in that measurement. 
The speed difference is measured at the time of peak brightness. The radiant difference is the angular separation between the computed radiants. As many of the radiant differences in our calibration set are quite large, with differences of up to $92^\circ$, we model the uncertainty of the log-radiant instead of the radiant.
We assume a Gaussian distribution for measurement uncertainty in speed and a log-Gaussian distribution for uncertainty in log-radiant. The uncertainty distributions are then empirically fit to the data.
The resulting estimated distributions for the speed and log-radiant uncertainties are shown in Fig. \ref{fig:uncertainty_histograms}.

\begin{figure*}[t]
    \centering
    \includegraphics[width=1.\textwidth]{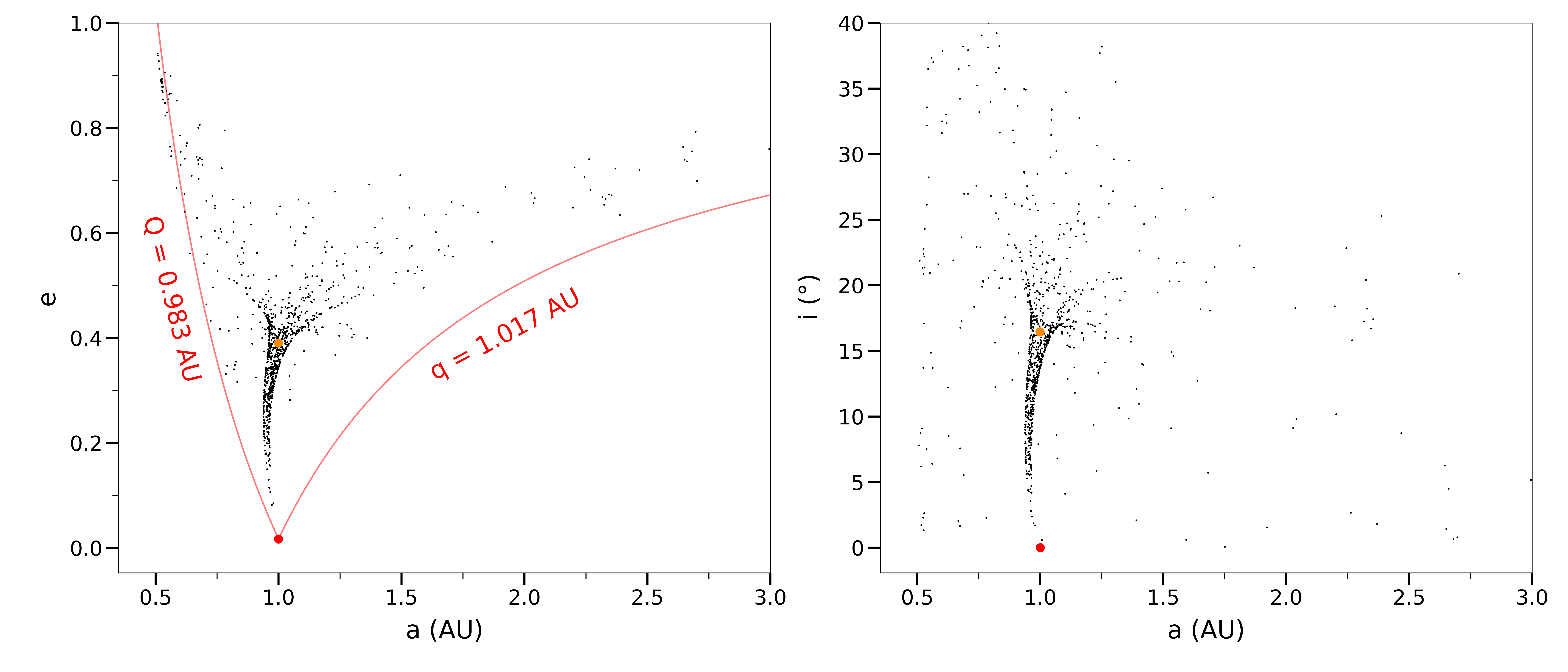}
    \caption{An example of the Monte Carlo cloning process for a single decameter impactor, the December 25, 2010 Pacific Ocean fireball. The object's nominal semi-major axis $a$ is plotted in orange against the eccentricity $e$ (left) and inclination $i$ (right), with orbits for each of the $1000$ Monte Carlo clones plotted in black.
    }
    \label{fig:mc_single_event}
\end{figure*}

Finally, clones of each decameter impactor are created by drawing Monte Carlo samples from the estimated speed and radiant uncertainty distributions and adding the sampled uncertainties to the respective impactor state vector. The pre-impact orbit of each clone is then computed from its modified state vector. Cloned orbits with an eccentricity greater than $e > 0.98$ or with a perihelion distance smaller than the radius of the sun ($q < R_\odot$) are rejected as unphysical. This process is then repeated until $1000$ clones have been generated for each impactor.
The results of this process for a single decameter impactor are shown in Fig. \ref{fig:mc_single_event}.  

The calculated $1\sigma$ orbital uncertainties for each impactor in Table $2$ show that the resulting clones have a large spread in orbital space, again demonstrating the large uncertainty in some USG measurements compared to other techniques. The orbital element most sensitive to uncertainty in this approach is the semi-major axis. In contrast, elements such as perihelion distance have much lower uncertainty, and the ascending nodal longitude is known very precisely (reflecting the fact that all impactors by definition collide with the Earth).

We note that previous studies assessing the general accuracy of USG sensor data \citep{devillepoix_observation_2019, brown_proposed_2023} have found the reported energy, location and height to be generally accurate, while the velocity and radiant are the least accurate measurements. As such, these are the primary variables that we focus on in our Monte Carlo uncertainty estimation method.

We have chosen to use this simpler approach of assuming Gaussian uncertainty described here due to the difficulty of accurately estimating a probability distribution from extremely limited number statistics of just $18$ objects. However, for completeness, we also conducted a reanalysis of all subsequent work presented in this paper using a nonparametric kernel density estimation approach to estimate the speed and radiant uncertainty distributions for generating the Monte Carlo clones, and found that it did not significantly change our conclusions. 

We also note that there appears to be a temporal trend in the uncertainties. As shown in \ref{app:usgtime}, the USG events with independently determined trajectories measured since 2018 have noticeably lower radiant/speed differences than earlier estimates. While we do not make use of this trend, it suggests that recent or current USG fireball orbits/trajectories may be more accurate than the historical dataset as a whole.

\section{Orbital similarity}\label{sec:orbital_similarity}

As decameter-sized NEOs are usually at the detection limit of most telescopic surveys, previous studies have suggested numerous observational biases for the telescopically observed NEO population \citep{jedicke_fast_2016, nesvorny_neomod_2023} in addition to very low detection efficiency overall \citep{harris_population_2021}. 
However, this result has never been quantitatively demonstrated due to the lack of an unbiased population to compare to the telescopic data.
Here we attempt to show this result for the first time by analyzing whether the observed telescopic decameter NEA population is representative of the true population (which we estimate using our decameter impactor sample) or if there are telescopic detection biases as previously suggested.

We perform this analysis by directly comparing the $3$-D distributions of the two populations in $\left(a, e, i\right)$ parameter space, using a multivariate version of the one-dimensional Kolmogorov-Smirnov (K--S) statistical test first proposed by \citet{peacock_two-dimensional_1983} and extended by \citet{fasano_multidimensional_1987}. The test evaluates the likelihood that two samples are drawn from the same underlying distribution.

We note that the kinetic energies of even the smallest decameter impactors are typically $\sim 7-8$ kT TNT, well above the detection limit of $\sim 0.1$ kT TNT for USG sensors. As such, we are confident that our sample of $14$ decameter impactors represents every decameter-sized object to strike Earth within the last $30$ years and we therefore consider it an unbiased sample of the true NEA population.

We also test for evidence of recent tidal disruption in the decameter population by comparing
the orbits of individual objects using the $D$ dissimilarity criterion, a metric formulated by \citet{southworth_statistics_1963} to associate meteors with known meteor showers and parent bodies, as well as to distinguish showers from the sporadic background. Several modifications of the $D$ criterion have since been developed \citep[e.g.][]{drummond_test_1981, jopek_remarks_1993, valsecchi_meteoroid_1999, jenniskens_meteoroid_2008}, each with different orbital parameter weightings.
Here we use the original $D$ criterion proposed by \citet{southworth_statistics_1963} to evaluate the tidal disruption hypothesis proposed to explain the excess meteoroid impact flux in the decameter size range.
In the case of a recent tidal disruption, occurring on a timescale less than the decoherence time for tidally-disrupted NEO fragments, which is of order 
$10^4$-$10^5$ 
years \citep{pauls_decoherence_2005, schunova-lilly_properties_2014, shober_near_2024}, we would expect strong orbital similarity between the impactors, as they would originate from the same parent body or bodies.

\subsection{Orbital parameter comparison}\label{sec:orbital_similarity_ks_test}

The multidimensional K--S test method of \citet{fasano_multidimensional_1987} computes the $n$-dimensional test statistic as the maximum test statistic for the cumulative distribution functions of the two samples when considering all $2^n$ possible point orderings in $\mathbb{R}^n$. In their original paper, \citet{fasano_multidimensional_1987} do not formulate an analytical method for significance testing, instead using Monte Carlo simulation to estimate critical values of the test statistic for two- and three-dimensional distributions. Here we adopt the permutation approach for significance testing of \citet{puritz_fasanofranceschinitest_2023} in their \texttt{R} implementation of the multidimensional K--S test.
Our Python implementation of the \citet{fasano_multidimensional_1987} multidimensional K--S test is available at \href{https://github.com/wmpg/fasano-franceschini-test}{github.com/wmpg/fasano-franceschini-test}. 

\begin{figure}[t]
    \centering
    \includegraphics[width=0.51\textwidth]{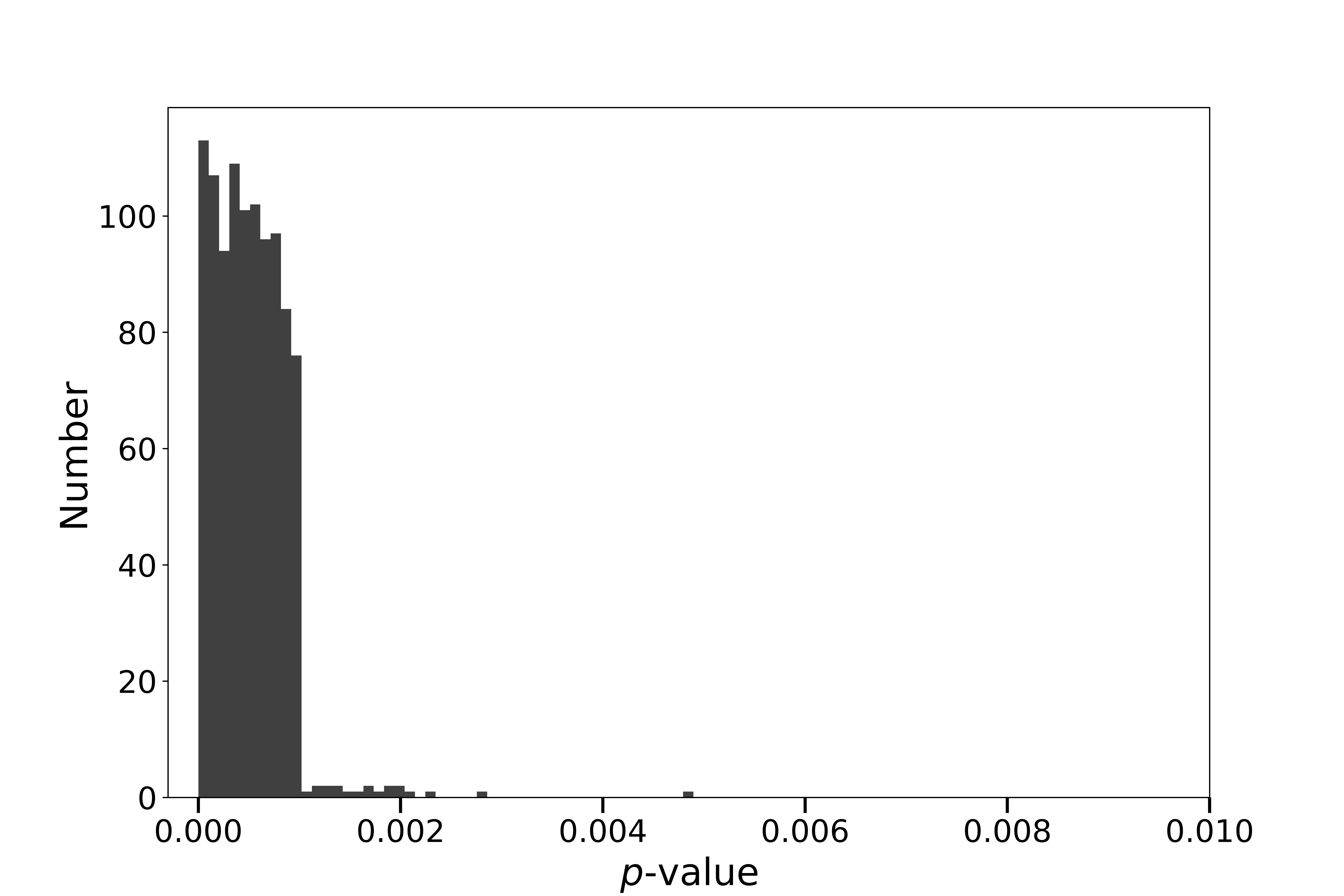}
    \caption{Distribution of $p$-values for the two-sample test statistic of the multidimensional Kolmogorov-Smirnov test 
    outlined by \citet{fasano_multidimensional_1987} 
    when comparing the $\left(a, e, i\right)$ distributions of the telescopic population against each of the $1000$ Monte Carlo cloned impactor populations.
    Most $p$-values are very small ($p \lesssim 10^{-3}$), with the largest $p$-value for any of the Monte Carlo samples still less than $0.01$. This represents strong evidence that the impactor and telescopically observed populations of decameter objects are not drawn from the same underlying $\left(a, e, i\right)$ distribution, and that the telescopically observed population is therefore biased in some way. 
    }
    \label{fig:p_value_ks_test}
\end{figure}

Using the multidimensional K--S test, we compared the test statistic of the $\left(a, e, i\right)$ distribution of the telescopic decameter NEA population against each of the $1000$ Monte Carlo samples of cloned impactors. The corresponding test statistics and $p$-values for each sample were then computed using the permutation test method of \citet{puritz_fasanofranceschinitest_2023} for $1000$ permutations. The distribution of $p$-values is shown in Fig. \ref{fig:p_value_ks_test}, with
most $p$-values being extremely small (on the order of $10^{-4}$) and even the largest $p$-value below $0.01$. 
We therefore find strong quantitative evidence that the decameter impactor and telescopically observed decameter NEA populations are not drawn from the same underlying $\left(a, e, i\right)$ distribution, as expected. This is likely due to the aforementioned telescopic biases and generally low detection efficiency, as well as the dependence of impact probability on orbital properties. 

\subsection{Short-term tidal disruption: Comparing $D$ dissimilarity criteria}\label{sec:orbital_similarity_d_criteria}

As an additional step in analyzing the decameter population, we wish to explore orbital similarity within the population on a per orbit basis. Any two objects showing unusually high orbital similarity could be an indication of a common origin, similar to the case found for asteroid pairs \citep{pravec_asteroid_2019} on a timescale of order 10$^4$ years. This is also roughly the timescale over which near-Earth orbits retain orbital coherence \citep{pauls_decoherence_2005, schunova-lilly_properties_2014, shober_near_2024}. High orbital similarity within the decameter population would therefore be evidence for a recent breakup event.

Here we use the $D$ dissimilarity criterion of \citet{southworth_statistics_1963} to evaluate orbital similarity, where
the function $D\left(x_1, x_2\right)$ quantifies the dissimilarity between two orbits $x_1$ and $x_2$. A lower $D$ value indicates that the orbits are more similar. 
The standard procedure for identifying meteor streams or determining a stream's association with a given parent body is to calculate the $D$-criterion $D\left(x_1, x_2\right)$ for two orbits $x_1$ and $x_2$. The orbits are considered similar if $D\left(x_1, x_2\right) < D_r$, where $D_r$ is a constant rejection threshold. This procedure is then repeated for all orbits in a given population. 

\begin{figure}
    \centering
    \includegraphics[width=0.51\textwidth]{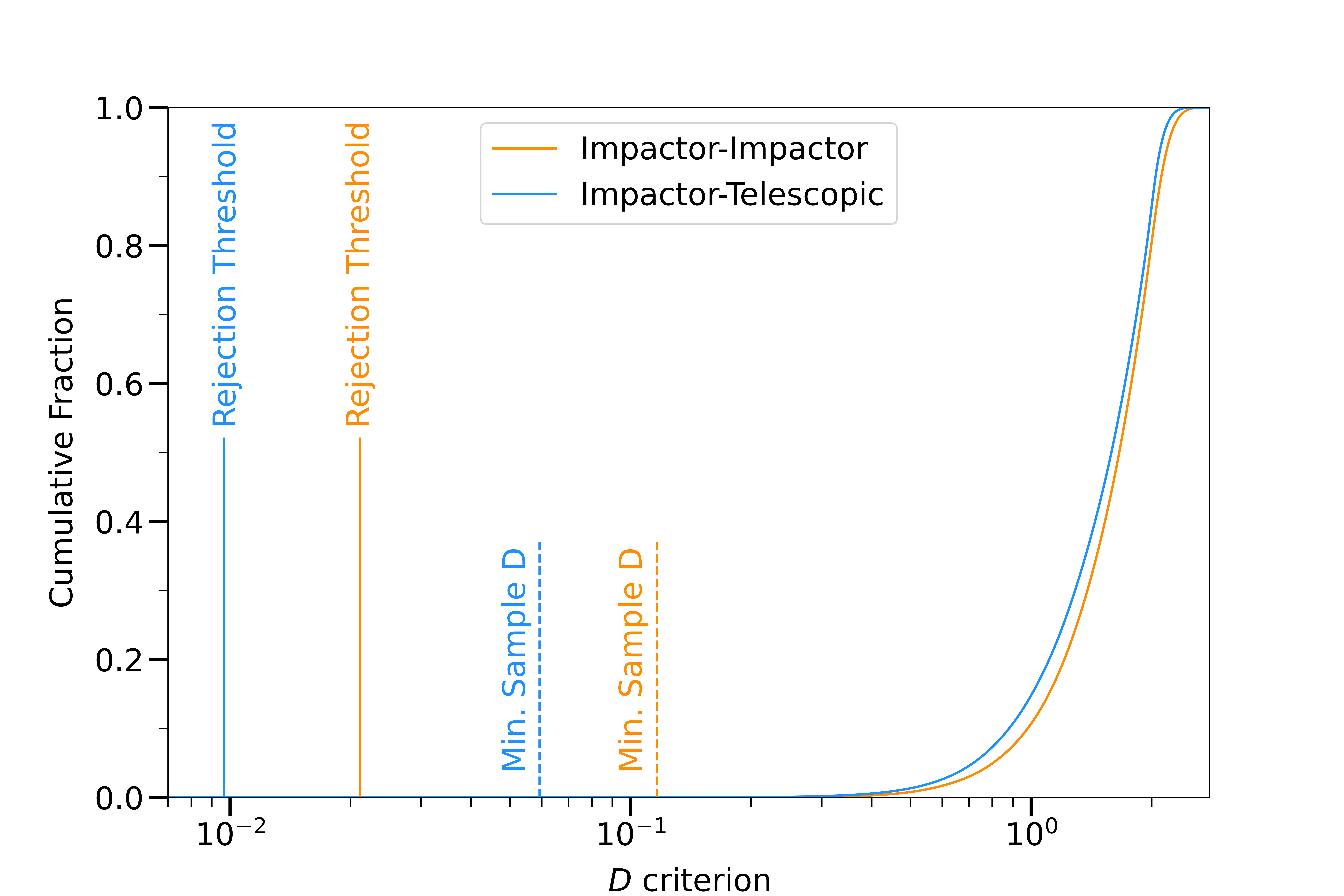}
    \caption{The cumulative distribution functions for the $D$ criteria of all cloned impactor orbits both compared to 
    all other impactors and their clones
    (orange) and to the telescopically observed population (blue). Note that the $x$-axis has been plotted on a log scale. The solid lines are the rejection thresholds $D_r$ (chosen as described in the text) above which two orbits are considered dissimilar, while the dotted lines mark the lowest $D$ criterion (most similar orbits) computed for any pair of orbits in the sample. 
    As the respective rejection thresholds of both samples are below even the smallest measured $D$ value (i.e. none of the orbits in either sample are considered similar), this suggests there is no common recent origin among all impactors/telescopic NEAs. This represents evidence against the theory that a very recent tidal disruption is responsible for the excess impact flux in the decameter size regime.}
    \label{fig:impactor_telescopic_cdfs}
\end{figure}

In their original formulation of the $D$ criterion, \citet{southworth_statistics_1963} calibrate the rejection threshold $D_r$ using a sample of $360$ meteors in a known shower, setting $D_r = 0.20$ as it is the highest $D$ value for any pair of meteors in their sample. 
\citet{southworth_statistics_1963} and \citet{lindblad_computerized_1971} suggest that the rejection threshold $D_r$ should depend approximately on the sample size $N$ as $D_r \propto N^{-1/4}$. This yields the following approximate rule for setting $D_r$, which we also use for the sake of simplicity in this work:
\begin{equation}
    D_r = 0.20 \times \left(\frac{360}{N}\right)^{1/4} \label{eqn:lindblad_threshold}
\end{equation}
We therefore start by computing the $D$-criterion $D\left(i_1, i_2\right)$ to compare the orbit of each decameter impactor and its Monte Carlo clones to those of every other impactor and their clones. We also compute $D\left(i, t\right)$ to compare each decameter impactor and all its clones to every telescopically observed decameter NEA. In both cases, we then compare
the calculated $D$-criteria to the respective sample rejection threshold $D_r$, which is rescaled based on the sample size according to Equation \ref{eqn:lindblad_threshold}. Fig. \ref{fig:impactor_telescopic_cdfs} shows the cumulative distribution functions of the $D$-criteria when comparing the impactor orbits and their clones both to each other and to the telescopically observed orbits.
The solid lines are the rejection thresholds $D_r$, while the dotted lines mark the lowest $D$ criterion (most similar orbits) computed for any pair of orbits in the population. In both cases, even the lowest $D$ criterion in the sample is greater than $D_r$, meaning that none of the orbits in either sample are considered similar. 
We conclude there is no evidence for recent physical association of any observed decameter impactor either with any other decameter impactor or with the telescopically observed decameter NEA population, suggesting that the recently observed decameter impactor population is not dominated by young tidal disruption fragments.

However, the small number of both decameter impactors and telescopically observed decameter NEAs and the short decoherence timescales of meteoroid streams produced by tidal disruptions mean that a more modest tidal disruption signature in the decameter population cannot be conclusively ruled out.
\citet{shober_near_2024} analyzed the $D$-criterion and other orbital similarity measures and found that near-Earth orbits exhibit chaotic behavior with characteristic timescales on the order of $10^3$ years, even shorter than the decoherence times of meteoroid streams.
Indeed, \citet{shober_near_2024} argue that associating observed fireballs to near-Earth meteoroid streams using orbital similarity measures like the $D$-criterion alone is not possible without a much larger dataset due to their rapid decoherence.
As such, we cannot strictly rule out a recent tidal disruption from orbital comparisons alone as the cause of the ``decameter gap" based on the limited available data, but merely note that no strong signature is evident.

\section{Long-term tidal disruption: Comparing perihelia distributions}\label{sec:long_term_td}

While we find no strong signature of recent ($\lesssim10^4$ years) tidal disruptions as the cause of the excess decameter impact flux, the possibility remains for tidal disruption to produce a population-level enhancement of the near-Earth asteroid or meteoroid environment from much older tidal disruption events. \citet{granvik_tidal_2024} show that NEA orbital distribution models based on delivery from the main-belt underpredict the number of telescopically observed NEAs with perihelia near the orbits of Venus and Earth. They argue that this excess is the result of tidal disruptions. They also show that an excess in perihelia near Venus and Earth is a direct consequence of tidal disruption by these bodies and that this pattern persists for long periods. Here we explore the possibility that such long-term tidal disruptions by Venus and Earth could be the cause of the decameter impact rate discrepancy as well and that such a signature would be present in the perihelia of decameter (and smaller) objects at Earth.

To evaluate the evidence for long-term tidal disruption, we compare four different data sets of Earth-impacting meteoroids, selecting only those of asteroidal origin, at distinct size regimes ranging from decameter- to millimeter-sizes. We then analyze the distribution of perihelion distances to determine whether they show enhancement over the steady-state delivery rate at the semi-major axes of Venus and Earth, as suggested by the analysis of \citet{granvik_tidal_2024} for telescopically observed NEAs. The first two datasets are CNEOS-derived orbits while the other two datasets are from ground-based optical cameras and used to provide context and comparison with the CNEOS population of larger bodies. 

The first data set consists of the $14$ decameter objects we identified earlier, and their associated Monte Carlo clones. We also use all reported fireballs in the CNEOS database which are smaller than decameter size. After filtering the entire CNEOS database and selecting only asteroidal meteoroids (employing the definition adopted by \citet{borovicka_data_2022}), we identify 
$249$
CNEOS fireballs smaller than decameter size for which pre-impact orbits can be determined. Assuming a similar bulk density of $\rho = 1500$ kg/m$^3$ as for the decameter objects, these objects range in size from $\sim1$-$7.5$ meters. We produce $1000$ Monte Carlo clones for each object using the same procedure as for the decameter impactors outlined in Section \ref{sec:uncertainty_estimation}. These 
$249$
meter-sized CNEOS fireballs, and their respective Monte Carlo clones, thus comprise our second data set.

We also explore a set of $824$ fireballs observed during $2017-2018$ by cameras of the European Fireball Network (EFN) for which orbits and orbital uncertainties have been computed by \citet{borovicka_data_2022-1}, ranging in size from $\sim2$-$40$ centimeters. After filtering out orbits with $e > 0.98$ or $q < R_\odot$ (as for the CNEOS data) and selecting only asteroidal meteoroids following the definition of \citet{borovicka_data_2022}, we obtain $323$ fireballs. We then generate $1000$ Monte Carlo clones for each object by drawing from a Gaussian distribution with variance equal to the square of the orbital uncertainties reported by \citet{borovicka_data_2022-1}. Our third dataset therefore consists of these $323$ centimeter- to decimeter-sized EFN fireballs and their corresponding clones.

Finally, our fourth data set consists of $101631$ meteors 
observed by Electron-Multiplied Charge-Coupled Device (EMCCD) cameras of the Canadian Automated Meteor Observatory \citep{gural_development_2022} operated by the University of Western Ontario. These EMCCD-observed meteors range from millimeter to centimeter sizes. We do not produce Monte Carlo clones for each of these EMCCD meteors due to computational cost, as well as the much larger sample size and generally smaller orbital uncertainties compared to the other three data sets.
This dataset has also been filtered to select only meteoroids which are likely asteroidal, again following the definition of \citet{borovicka_data_2022}.

For each of these four datasets, we examine the distribution of impactor perihelion distances, $q$, weighted by inverse collision probability with the Earth. The annual collision probability of each meteoroid (and Monte Carlo clone, when applicable) is computed using the approach described by \citet{pokorny_opik-type_2013}, due to its improved accuracy for high-inclination and high-eccentricity orbits compared to previous methods \citep[e.g.][]{opik_collision_1951, wetherill_collisions_1967}. All four sets of data are then individually binned by perihelion distance $q$ into bins of size $0.1$ AU. This size is chosen to match the binning used for the orbital distribution models of \citet{granvik_super-catastrophic_2016} and \citet{granvik_debiased_2018}, allowing for direct comparison. 

In each bin, we compute the $2.5\%$ and $97.5\%$ percentiles for impact probability, and then discard points lying outside this range to reduce the influence of extreme outliers on the weighted distributions. We examined these distributions choosing $0.5\%$ and $99.5\%$ filter limits as well and found no significant difference.
Each meteoroid is then weighted by the inverse of its collision probability (i.e. objects less likely to collide with the Earth are weighted more strongly) when computing the distribution of perihelion distances.

Finally, the uncertainty for each bin is then estimated by bootstrapping the original sample, a procedure by which the sampling distribution of an estimator can be inferred by resampling the original data
\citep{efron_bootstrap_1979, davison_bootstrap_1997}. Here we draw $1000$ bootstrap samples with replacement from all four data sets, each with a sample size equal to that of the original data set. For the three data sets where Monte Carlo clones have been generated, the associated Monte Carlo clones for each selected event are included in the bootstrap sample as well. Each bootstrap sample is then binned by $q$ into bins of width $0.1$ AU, and outliers are identified and discarded using the same method described earlier. The weighted distribution of each bootstrap sample is then computed, and the error bars for each bin are taken to be the $1\sigma$ standard deviation in that bin's weighted value across all $1000$ bootstrap samples. 

\begin{figure*}
    \centering
    \includegraphics[width=1.\textwidth]{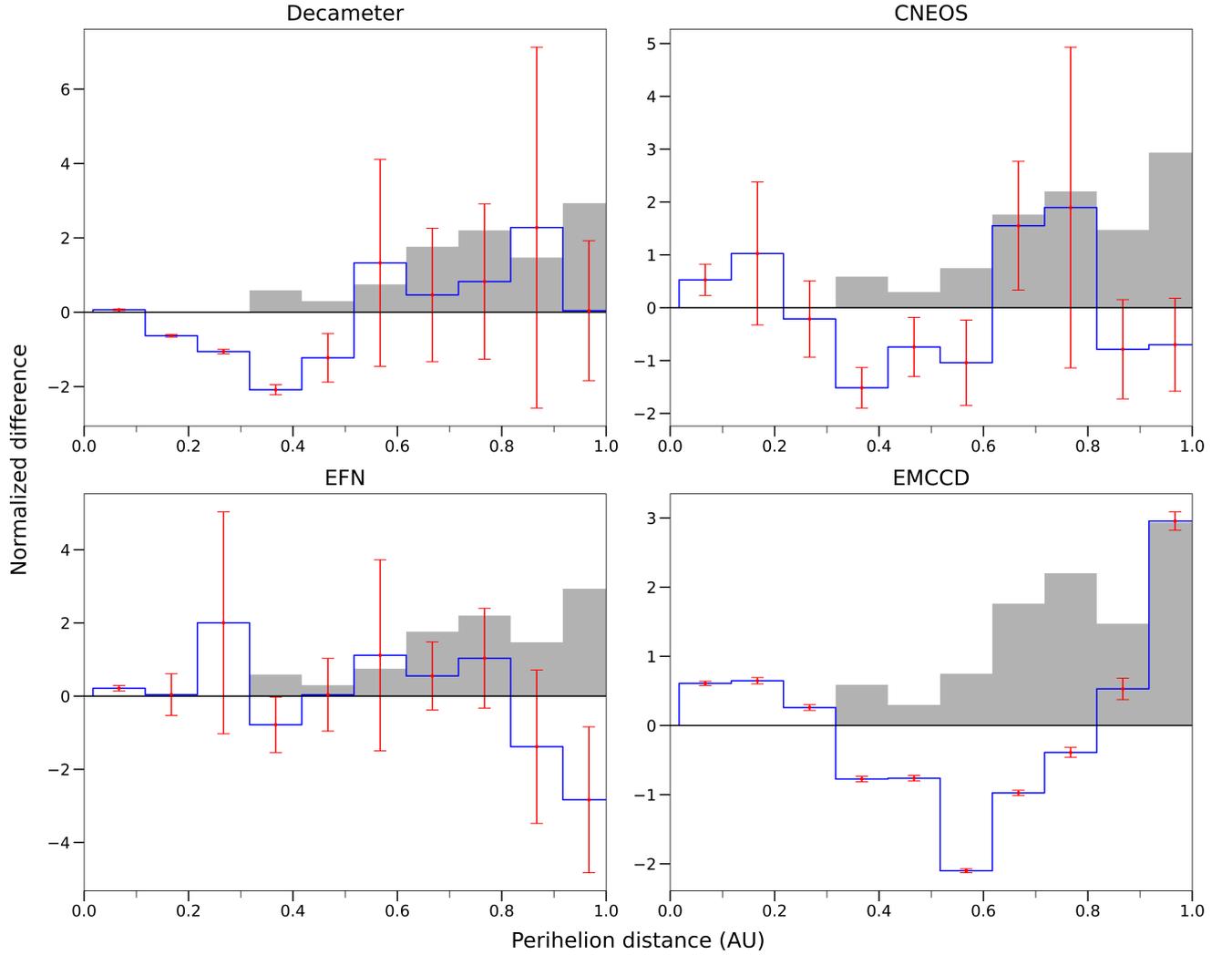}
    \caption{The perihelion distribution for four different sets of Earth-impacting meteoroids at distinct size regimes: decameter-sized (upper left), meter-sized (upper right), centimeter/decimeter-sized (lower left) and millimeter-sized (lower right).
    The blue line shows the difference between the observed $q$ distribution of meteoroids weighted by annual impact probability, and the predicted NEO $q$ distribution by the model of \citet{granvik_super-catastrophic_2016}. The gray histogram is the $q$ distribution of synthetic objects undergoing B-type tidal disruptions in numerical simulations, as reported in Fig. 1 of \citet{granvik_tidal_2024}. Both histograms have been normalized to a total absolute area of unity. The red error bars were found by bootstrapping each dataset as described in the text and represent $1\sigma$ uncertainties. For the CNEOS, EFN and EMCCD datasets, only those meteoroids which are asteroidal according to the definition of \citet{borovicka_data_2022} are included in the analysis.
   }
    \label{fig:q_histograms}
\end{figure*}

The bin size of $0.1$ AU is larger than the spread in the CNEOS and decameter perihelia variance found using our cloning procedure. We found the CNEOS clones had a mean difference of $0.02$ AU and standard deviation of $0.010$ AU in their perihelion distance; hence, we conclude that very little blurring across bins occurs amongst clones from a single fireball. Indeed, we also find that using the perihelion distribution of only the observed CNEOS events (without cloning) produces a histogram within the uncertainties shown in the upper right sub-plot of Fig. \ref{fig:q_histograms}.

Next, we compared the distribution of perihelion distances for our four observed sets of meteoroids (decameter, CNEOS, EFN and EMCCD) to the distribution of NEO perihelion distances predicted by the orbital distribution model of \citet{granvik_super-catastrophic_2016}. That model is based on a catalog of $3,632$ distinct NEOs observed by the Catalina Sky Survey (CSS) with absolute magnitudes ranging from $17$ to $25$ during the period 2005-2012. 
Our results are summarized in Fig. \ref{fig:q_histograms}. The blue line shows the difference between the observed $q$ distribution of asteroidal orbital-type meteoroids for the four sets of data considered and the predicted NEO $q$ distribution according to the model of \citet{granvik_super-catastrophic_2016}, normalized to a total absolute area of unity. The associated uncertainties estimated by bootstrapping are shown by the error bars. The gray histogram shows the $q$ distribution (also normalized to unity area) of simulated objects undergoing B-type tidal disruptions during encounters with Earth or Venus as shown in Fig. 1 of \citet{granvik_tidal_2024}, for comparison.

The decameter distribution shows large uncertainties for each bin reflecting the very low number statistics in that sample. We cannot conclude anything for this population. The CNEOS sample similarly has large uncertainties per bin. Only bins near $0.6$-$0.8$ AU show a slight enhancement and these are of order 1$\sigma$ deviations. 
We conclude that our measurements therefore do not confirm a tidal excess in current perihelia compared to the steady-state NEA model, though we cannot strongly rule this out as our statistics are much lower than those presented for larger NEAs by \citet{granvik_tidal_2024}.

The EFN dataset (centimeter--decimeter sizes) shows no excess for meteoroids selected according to the asteroidal orbit definition of \citet{borovicka_data_2022}. Interpretation of the EMCCD distribution is more complex. While the nominal \citet{borovicka_data_2022} definition appears to produce an excess near the Earth, a slight increase in the $T_J$ threshold for asteroidal meteoroids from $3.0$ from $3.2$ removes this feature. \citet{tancredi_criterion_2014} notes that some mixing of asteroidal and cometary bodies occurs near $T_J\sim3$, so this change may simply reflect cometary contamination of the population, particularly from comet 2P/Encke \citep{wiegert_dynamical_2009}. A similar definition change does not alter the EFN distribution as much. For the EMCCD population, a further contributing factor may be the small size of the meteoroids in the dataset, which are more strongly affected by radiation forces and hence less likely to follow strictly gravitational evolution from the main-belt to Earth-crossing orbits. 

While there is a deficit in the $0.4-0.6$ AU range of the CNEOS and EMCCD samples, this is not a signal of tidal disruption. As our intent in this paper is specifically to evaluate the possibility of tidal disruption by looking for enhancement near the semi-major axes of Venus and Earth, we do not focus on other features in the histograms.

We conclude that there is no evidence for long-term tidal disruption in the currently observed Earth-impactor population. Of the four datasets we analyze, three do not show any tidal disruption signatures at all. Only the $q$ distribution of the CNEOS data shows a $\sim1\sigma$ enhancement in the 0.6-0.8 AU range, which we suggest is most likely a statistical anomaly.

\section{Source regions}\label{sec:source_regions}

We are also able to estimate where both the impactor and telescopically observed populations of decameter-sized objects originate from in the main asteroid belt. This is accomplished using the four-dimensional model for NEO distribution in 
$a$, $e$, $i$, and $H$
described by \citet{granvik_debiased_2018}. Given an object's $a$, $e$, $i$ and $H$, the model of \citet{granvik_debiased_2018} computes the probabilities of the object originating from one of seven different source regions (also called escape regions) in the main asteroid belt. These source regions are the $\nu_6$ secular resonance with Saturn, the 3:1, 5:2 and 2:1 mean-motion resonances with Jupiter, the Hungaria and Phocaea asteroid families, and the Jupiter-family comets.

As the \citet{granvik_debiased_2018} model distribution is only sampled over a $4$-D rectilinear grid with a step size of $0.1$ in $a$, $0.04$ in $e$, $4^\circ$ in $i$, and $0.250$ in $H$, intermediate points in the parameter space are interpolated using the tensor product of one-dimensional cubic spline interpolations for each parameter. Data points falling outside of the sampled region are discarded in our analysis.
Moreover, the \citet{granvik_debiased_2018} model distribution is only sampled up to a maximum absolute magnitude $H = 25$, below even the brightest decameter objects we consider. As such, we manually assign all objects in both our impactor and telescopic populations an absolute magnitude of $H = 25$ to ensure that they fall within the sampled region of the distribution. 
While the \citet{granvik_debiased_2018} model may not be entirely accurate for these fainter decameter objects as they lie outside the sampled region of parameter space, this is the best estimate available given the lack of NEO distribution data for objects of this size.

\begin{figure*}[t]
    \centering
    \includegraphics[width=1.\textwidth]{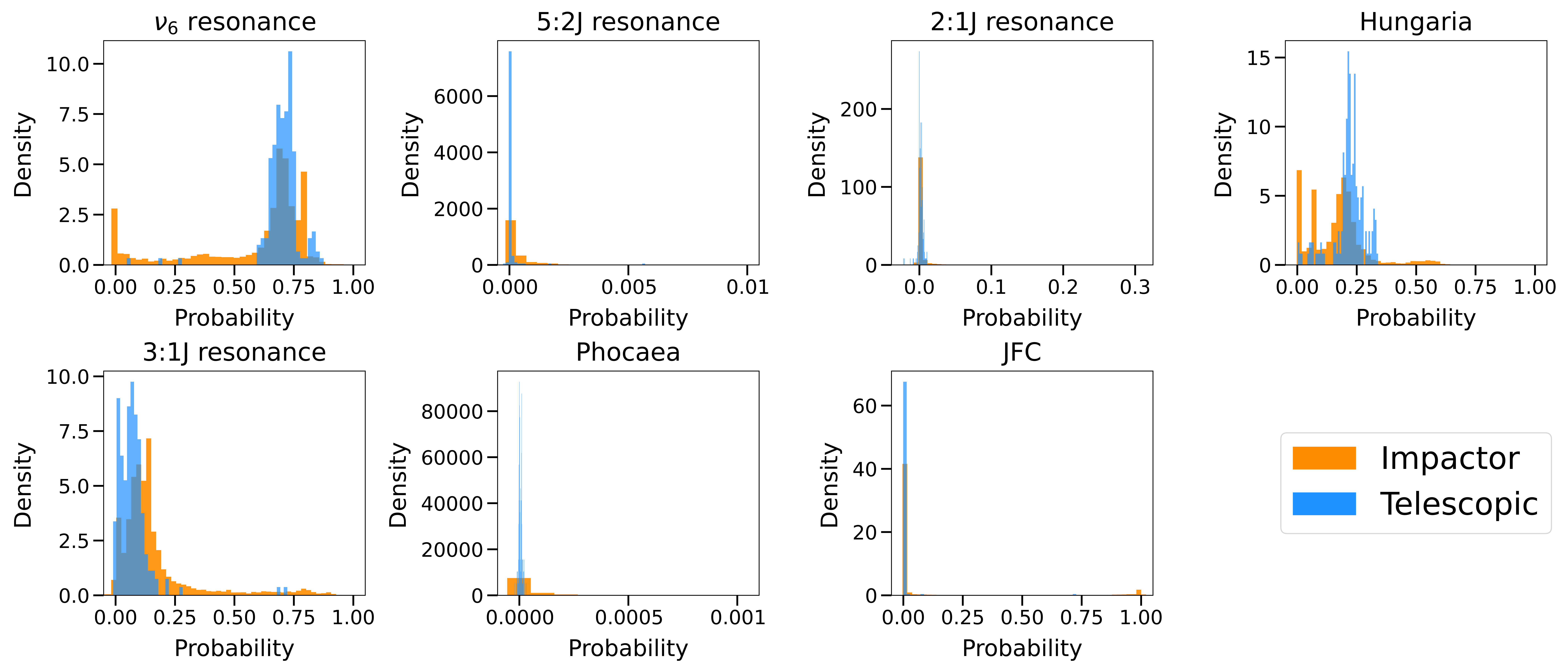}
    \caption{Normalized histograms of probabilities computed for each of the seven source regions considered in the model of \citet{granvik_debiased_2018}, for the cloned decameter impactor (orange) and telescopically observed (blue) populations. Note that the abscissa are scaled differently for each source region.}
    \label{fig:source_regions}
\end{figure*}

We apply the \citet{granvik_debiased_2018} model both to our telescopically observed decameter NEA population as well as the $1000$ Monte Carlo cloned impactors, to probabilistically determine their source regions. The results are shown in Fig. \ref{fig:source_regions}.

To zeroth order, both the decameter impactor and telescopic decameter NEA populations appear to broadly originate from the same source regions: primarily from the $\nu_6$ secular resonance ($\sim70\%$ of objects) with smaller contributions from the Hungaria group ($\sim20\%$) and the 3:1 Jupiter mean-motion resonance ($\sim10\%$). The contribution of the other four source regions to the decameter NEA population is negligible. This is consistent with the primary delivery escape regions for meteorites from the main-belt \citep{granvik_identification_2018} and similar to the source regions associated with NEOs having spectral affinities matching ordinary chondrites \citep{binzel_compositional_2019}.

\section{Discussion}\label{sec:discussion}

We find the decameter impactor population as a whole to have a different orbital distribution than the telescopic decameter population. This is unsurprising as the latter are expected to be heavily shaped by observational biases \citep{jedicke_fast_2016, nesvorny_neomod_2023}.
However, despite these population level differences, when using the \citet{granvik_debiased_2018} model to map the two populations back to likely main belt source regions, a similar picture for both emerges: namely, that most decameter impactors/telescopic NEAs originate from the $\nu_6$ resonance with minor contributions from the Hungaria and the 3:1 mean-motion resonance. We find no significant signal from any other source region.

This suggests that the present decameter impactor population is dominated by inner-mid main belt sources while outer belt delivery is negligible. We note that the telescopic population may be underestimating lower albedo objects with could have an outer belt source, though none of the decameter impactors have pre-impact orbits consistent with a purely outer belt source either.

This finding is broadly consistent with what is known of the dominant delivery escape regions in the main belt for ordinary chondrite meteorites \citep{granvik_identification_2018} and NEAs \citep{demeo_solar_2014, binzel_compositional_2019}. The lack of an outer belt source for impactors does not preclude weaker C-type asteroidal bodies amongst the impactor population, as these are almost as abundant from inner belt source regions as S-type asteroids when the telescopic NEA population is debiased \citep{demeo_solar_2014, marsset_debiased_2022}.

Among our decameter impactor population only one impactor has a known composition: Chelyabinsk, an LL chondrite. It has been shown that among all NEAs where spectral affinity with meteorite analogues has been estimated, LL-type NEAs are the most abundant sub-type \citep{vernazza_compositional_2008, sanchez_population_2024}. 
While the source region distribution of our impactor population is consistent with the five known LL-chondrite meteorite orbits in that they show a dominant $\nu_6$ and 3:1 signature \citep{brown_golden_2023}, the statistics are not strong. Indeed, \citet{broz_source_2024} and \citet{marsset_massalia_2024} have suggested that the Flora collisional family is instead the primary source of LL-chondrites.

Our results in section \ref{sec:orbital_similarity_d_criteria} suggest that the excess impactor flux is not due to recent tidal disruption. This conclusion is valid at centimeter-sizes where number statistics are reasonable and the effects of radiation forces (which may distort the $q$ distribution at EMCCD sizes) are negligible. 
However, at larger sizes we are limited by very small number statistics. While this is most true for the decameter impactors, at meter-sizes the USG impactor population is still more than an order of magnitude less in number than the NEA population used by \citet{granvik_tidal_2024} as evidence for tidal disruption.

One way to explain the decameter gap would be that impactors have systematically higher impact probabilities than the $f_\mathrm{imp} = 2.6 \times 10^{-9}$ yr$^{-1}$ average assumed in the NEA delivery model of \citet{nesvorny_neomod_2023}. Indeed, tidal disruption models predict that the fragments of such disruptions will be on more Earth-like orbits and have a higher chance of colliding with Earth. Hence, if the observed decameter population has a significantly higher apparent impact probability ($f_\mathrm{imp}$) than is usually assumed in converting the telescopic population to impactor numbers, this could explain the gap. Examining Table \ref{tab:nominal_orbits}, we see that the median impact probability for the nominal orbits of the $14$ decameter impactors is a factor of two higher than the canonically assumed NEA $f_\mathrm{imp}$ of $2.6 \times 10^{-9}$ yr$^{-1}$.

\begin{figure*}[t]
    \centering
    \includegraphics[width=1.\textwidth]{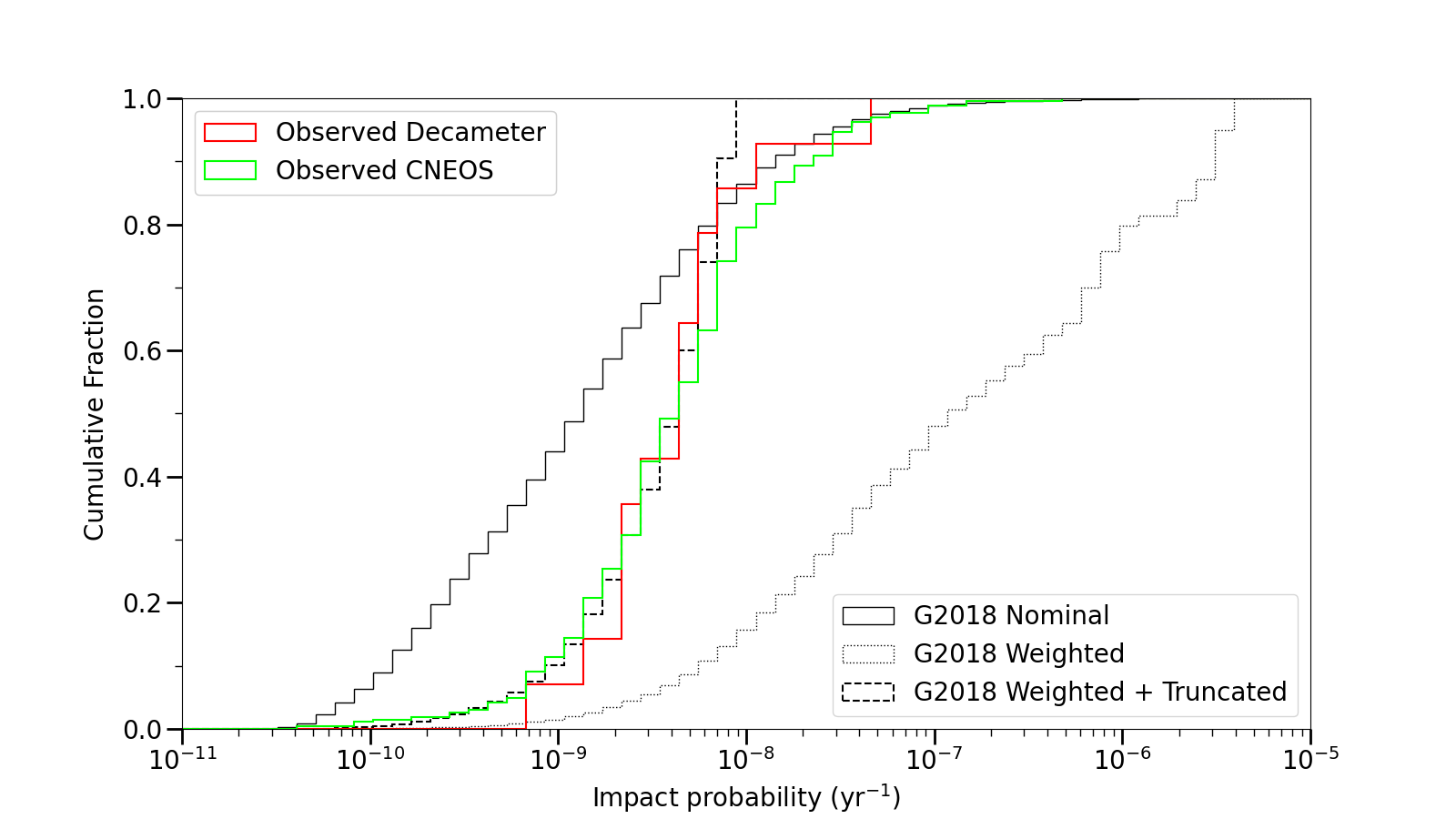}
    \caption{
    Cumulative distribution functions for the impact probabilities $f_\mathrm{imp}$ of observed decameter impactors (red) and of all observed CNEOS fireballs (green), compared to the impact probabilities of a synthetic decameter NEO population generated using the model of \citet[][black]{granvik_debiased_2018}, here denoted G2018. The solid black line shows the nominal $\mathrm{f_\mathrm{imp}}$ distribution of the entire NEO population. The dotted black line is weighted by $f_\mathrm{imp}$ and thus shows the expected $f_\mathrm{imp}$ distribution for the entire \emph{Earth-impacting} NEO population. The dashed black line is again weighted by $f_\mathrm{imp}$, and shows the expected $f_\mathrm{imp}$ distribution of Earth-impactors when the NEO population is limited to objects with nominal impact probabilities $f_\mathrm{imp} < 10^{-8}$ yr$^{-1}$. The latter distribution
    closely matches those of the observed decameter and CNEOS impactor populations. We suggest this is because objects with very large impact probabilities are typically scattered on timescales much shorter than their impact timescales.
    }
    \label{fig:IP}
\end{figure*}

However, we expect the observed impactor population to have a somewhat higher $f_\mathrm{imp}$ than the raw, unbiased population, as the impactor population is found by weighting each object in the raw population by its $f_\mathrm{imp}$. Figure \ref{fig:IP} shows the nominal $f_\mathrm{imp}$ cumulative distribution functions (CDFs) for the synthetic decameter population of \citet{granvik_debiased_2018}, as well as the expected impactor distribution derived from that CDF (weighted by $f_\mathrm{imp}$). The weighted impact probabilities predict a much higher fraction of large $f_\mathrm{imp}$ values than the observed decameter population. By limiting the model impact probabilities to $f_\mathrm{imp} < 10^{-8}$ yr$^{-1}$, we are able to reproduce all but the last $10\%$ of the observed decameter impact probability CDF, which is also very similar to that of the CNEOS population as a whole.

We suggest this indicates that the large model impact probabilities are not represented in the observed population as such large $f_\mathrm{imp}$ values (usually associated with NEAs at very low inclination or with $q$ or $Q$ near the Earth) are scattered on timescales much shorter than typical impact intervals. Though this does not preclude tidally-disrupted fragments from forming part of the impactor population, we do not see evidence for an unusually high $f_\mathrm{imp}$ distribution in the NEA population colliding with the Earth. On the contrary, the observed population is most consistent with a truncated (lower $f_\mathrm{imp}$) population than model predictions which nominally extend to much larger impact probabilities.

If future observations improve statistics enough to show that the meter-size (but not decameter-size) NEAs have an excess population and that this is due to past tidal disruptions, it could explain the disparity in the LL-chondrite falls vs. NEAs. In this scenario a past tidal disruption was from an LL-chondrite precursor. The lack of decameter-sized objects could potentially be explained if the precursor were comparatively small (only a few decameters in size itself), but this would not be consistent with the excess at larger sizes noted by \citet{granvik_tidal_2024} nor explain the decameter gap.
If both the meter- and decameter-size populations were shown to have an excess, it would be a natural explanation for the decameter gap, which in this scenario would extend to even smaller sizes and be consistent with the NEA excess of \citet{granvik_tidal_2024}. However, it would not address the discrepancy between the LL-chondrite meteorite flux compared to the NEA LL-chondrite population.

All of these interpretations remain speculative. Only more impactor observations at these sizes can reduce the large uncertainties to determine if the small excess we find for USG perihelion numbers in the $0.6$-$0.8$ AU range is anything other than a statistical anomaly, which we suggest is the most likely interpretation. 

\section{Conclusions}\label{sec:conclusions}

In this paper we 
presented the first population-level study of the dynamical properties of decameter-sized Earth impactors using energies and hence masses (and sizes), based on previously classified USG sensor fireball lightcurve data released in 2022 by the U.S. Space Force. We confirmed a significant discrepancy in the decameter size regime between the observed impact rate based on USG data as well as previous fireball estimates, and the impact rate of objects inferred from recent telescopic surveys of NEOs.
We analyzed the orbital properties of the decameter impactor and telescopic decameter NEA populations, particularly to evaluate tidal disruption as a possible explanation for this ``decameter gap."
Our key conclusions are summarized as follows:

\begin{enumerate}
    \item The decameter impactor and telescopic NEA populations are unlikely to be drawn from the same $\left(a, e, i\right)$ orbital distribution, suggesting a bias in telescopic detections. This is consistent with previous studies suggesting detections of small NEAs are likely to be affected by observational biases \citep{jedicke_fast_2016, nesvorny_neomod_2023}. Specifically, we found that objects from the impactor population generally had higher eccentricities and larger semi-major axes than those from the telescopically observed NEA population. 
    \item There is no evidence for recent ($\lesssim 10^4$ years) tidal disruption as the cause for the decameter impact rate discrepancy based on direct comparison of individual orbits. 
    \item Comparing the perihelion distribution among four sets of asteroidal meteoroids ranging from millimeter- to decameter-sizes, we find no significant excess suggestive of long-term tidal disruption. Moreover, the limited number of observations of meter-sized and larger impactors preclude any strong conclusions about long-term tidal disruption, and we argue that any observed deviation from the predicted distributions is likely a statistical anomaly. More observations are required to conclusively evaluate long-term tidal disruption as the cause of the decameter impact rate discrepancy.
    \item The decameter impactor and telescopic decameter NEA populations broadly originate from the same source regions of the main asteroid belt, mostly coming from the $\nu_6$ secular resonance ($\sim70\%$ of objects) with smaller contributions from the Hungaria group ($\sim20\%$) and the 3:1 mean-motion resonance with Jupiter ($\sim10\%$). The contributions of other source regions to both populations are negligible. 
    \item The observed decameter (and CNEOS) population have an impact probability distribution which matches that expected from the model of \cite{granvik_debiased_2018} if only synthetic NEAs with impact probability $f_\mathrm{imp} < 10^{-8}$ yr$^{-1}$ are included. This suggests that model-generated high-impact probability objects may not be resident long enough to affect the observed impactor population. No obvious impact probability excess attributable to tidal disruption is apparent.
\end{enumerate}
Our conclusions for the decameter impactor population are generally tentative due to small number statistics, and will require more observations to definitively confirm.
A future paper will characterize the physical and material properties of the decameter impactor population by modeling their ablation in the atmosphere using the newly available USG lightcurves.

\section{Acknowledgements}\label{sec:acknowledgements}
Funding for this work was provided in part by the Meteoroid Environment Office of NASA through co-operative agreement 80NSSC24M0060, the Natural Sciences and Engineering Research Council of Canada and the Canada Research Chairs program. We thank Mikael Granvik and an anonymous reviewer for providing feedback on an earlier version of this manuscript and Mikael Granvik for providing the original CSS input data used in his NEO model. Bill Bottke, David Nesvorn\'y, David Vokrouhlick\'y and Althea Moorhead provided many helpful discussions. 



\appendix

\section{USG uncertainties over time} \label{app:usgtime}
\begin{figure}[t]
    \centering
    \includegraphics[width=0.51\textwidth]{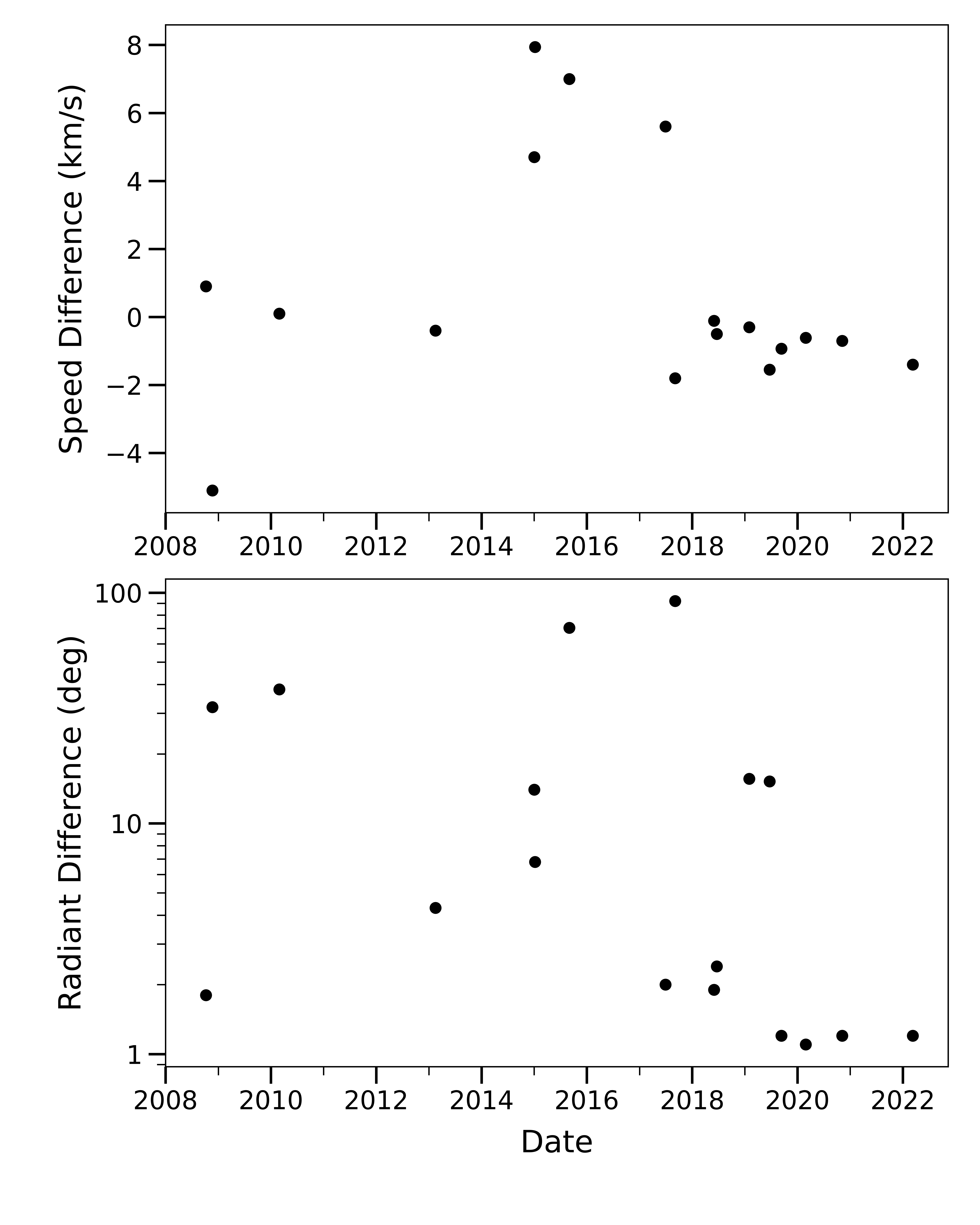}
    \caption{The speed and radiant difference between USG measurements and independently determined trajectories for the $18$ calibration fireballs used to estimate uncertainties in Section \ref{sec:uncertainty_estimation} are plotted by impact date. Note that the radiant is a log scale. Recent events (observed since $2018$) display much smaller differences than earlier events.}
    \label{fig:usgtime}
\end{figure}

As discussed in Section \ref{sec:uncertainty_estimation}, the more recently recorded USG events display lower speed/radiant differences compared to independent trajectory estimates. This is shown in Fig. \ref{fig:usgtime}.

\bibliographystyle{elsarticle-harv} 
\bibliography{Decameter_paper}





\end{document}